\documentclass[12pt,showpacs,preprintnumbers,nofootinbib,preprint]{revtex4}%
\usepackage{graphics}
\usepackage{graphicx}
\usepackage{pstricks}

\newcommand{\be}{\begin{eqnarray}}
\newcommand{\beq}{\begin{equation}}
\newcommand{\eeq}{\end{equation}}
\newcommand{\ee}{\end{eqnarray}}

\newcommand{\bmp}{\noindent\begin{minipage}{16cm}}
\newcommand{\emp}{\end{minipage}\vskip 7mm} 

\newcommand{\lsim} {\buildrel < \over {_\sim}}
\newcommand{\gsim} {\buildrel > \over {_\sim}}
\usepackage{bm}
\usepackage{amsmath}
\usepackage{amsfonts}
\usepackage{bbm}

\begin{document}
\title{Charge Asymmetry of Weak Boson Production at the LHC\\
and the Charm Content of the Proton}

\author{Francis Halzen}
\affiliation{Wisconsin IceCube Particle Astrophysics Center and Department of Physics, University of Wisconsin, Madison, WI, 53706}

\author{Yu Seon Jeong and C. S. Kim\footnote{cskim@yonsei.ac.kr}}
\affiliation{Department of Physics and IPAP, Yonsei University, Seoul 120-749, Korea}

\begin{abstract}
\noindent We investigate the production of the weak bosons $W^+$, $W^-$ and $Z$ at the LHC as a function of their rapidity, reconstructed {\it experimentally} from their leptonic decay. We show that the measurements provide a powerful tool for constraining the parton distribution functions, as was already the case for the lower energy $p\bar{p}$ collider data. We study the charge asymmetry determining the $u$ and $d$ distribution functions using the reconstructed $W^+$, $W^-$ rapidities. We also show how the ratio of the $W$ and $Z$ boson rapidity distributions directly probes the charm quark distribution function of the proton.
\end{abstract}

\maketitle

\newpage

\section{Introduction}

The weak bosons $W^+$, $W^-$ and $Z$ are produced in a hadron collider predominantly through the interaction of a quark and antiquark from each hadron, the so-called Drell-Yan mechanism \cite{DY}.
This process has been studied extensively as a test the Standard Model (SM) \cite{CDFg92, CDFr94, CDFc96, CDFW, CDFM,  D0c99, D0w00, D0W, D0M, CDFII05, CDFII07, D0II08, D0II09, CMS11, ATLAS10, ATLAS12}.

The cross sections for $W^\pm,~Z$ production have been computed in quantum chromodynamics (QCD) up to the next-to-next-to leading order (NNLO) \cite{Hamberg,VRAP} and provide a test of higher order QCD corrections.
The measured cross sections are related to the partonic QCD cross sections by the parton distribution functions (PDFs).
The partonic cross sections and the PDFs factorize making a direct study of the PDFs possible.

The PDFs can be examined using the rapidity dependence of the $W$ and $Z$ production cross sections.
Particularly, the asymmetry of the $W^+$ and $W^-$ rapidity distributions are mainly
from the difference of the $u(x)$ and $d(x)$ distributions, therefore it can determine $u(x)/d(x)$.
Since this method to probe the $u(x), ~d(x)$ using $W$ asymmetry was suggested
for lower energy $p\bar{p}$ collider \cite{BHKW},
it has been extensively investigated at both the Tevatron and the LHC
\cite{BHKW, Bodek, CDF09day, CFG, LFL, BAMO}.
Additionally a prescription was proposed to determine the charm quark structure function from both $W$ and $Z$ rapidity distributions measured at the Tevatron \cite{HHK},
where the charm quark contribution to $W,~Z$ production is small.
Using a similar method, the strange quark distribution has been studied at the LHC \cite{Kusina, ATL12s}.

In this paper, we investigate the PDFs by applying the methods in Refs. \cite{BHKW,HHK}
to $W$ and $Z$ rapidity distributions at the LHC center of mass energy $\sqrt{s}$ = 7 TeV,
to be increased to 14 TeV in the future.
Specifically, we evaluate the $W$ charge asymmetry with the $W$ rapidity,
which can be experimentally reconstructed from the momentum of the decayed leptons and neutrino;
we will discuss the details of the reconstructed rapidity later.
In addition to the $W$ charge asymmetry, we explore for the first time
the charm distribution function at the LHC with the reconstructed rapidity.
The LHC energies are significantly higher than the Tevatron energy of $\sqrt{s}$ = 1.8 TeV for Run I and 1.96 for Run II.
The higher energy not only leads to increased production rates, it provides access to the lower fractional momenta of the partons where sea quark interactions dominate; these were difficult to probe at Tevatron energies.
Thus, we expect to obtain the clear information especially of the charm distributions at the LHC.
In addition to the increased energy, the LHC collides protons, a symmetric process, resulting in a different pattern of rapidity distributions and charge asymmetry compared to $p\bar{p}$ collisions.
Thus, using the charge asymmetry and the ratio of the $W$ and $Z$ boson rapidity distributions,
we will eventually show how the LHC data, unlike the lower energy Tevatron data, directly probe $u(x)/d(x)$ and the charm quark distribution function, respectively.

This paper is organized as follows. In the following section, we introduce the schematic formalism
and show the rapidity distributions of $W^\pm$ and $Z$ bosons at the full NNLO in QCD.
In Section III, we present the method to reconstruct the $W^+$ and $W^-$ rapidity experimentally,
and investigate the charge asymmetry of the reconstructed rapidity distributions
to probe the $u(x)$, $d(x)$ PDFs.
In Section IV, we demonstrate the sensitivity of the charm quark PDF
to the ratio of the rapidity dependence of the $W$ and $Z$ cross section.
Finally, in Section V, we summarize our results.

\section{Rapidity Distributions of the differential cross sections}

The general expression for the cross sections for $W, Z$ production is given by
\be
\sigma_{AB \rightarrow W/Z} = \sum_{a,b} \int dx_a \int dx_b f_{a/A}(x_a, Q^2) f_{b/B}(x_b, Q^2)
\hat{\sigma}_{ab \rightarrow W/Z} \ ,
\label{eq:xsec}
\ee
where the partonic scattering cross section is $\hat{\sigma}_{ab \rightarrow W/Z}$.
In Eq. \ref{eq:xsec}, $x_{a(b)}$ are the momentum fractions carried by the partons $a(b)$ in the colliding hadrons $A(B)$, and $f_{a/A}(x_a,Q^2)$ and $f_{b/B}(x_b, Q^2)$ are the PDFs of $a$ and $b$, respectively.
The scale $Q$ is referred to as the factorization scale, $\mu_F$,
and it is set to be the mass of the produced boson, $\mu_F = M_{W/Z}$.

\begin{figure}[h]
\includegraphics[scale = 0.23]{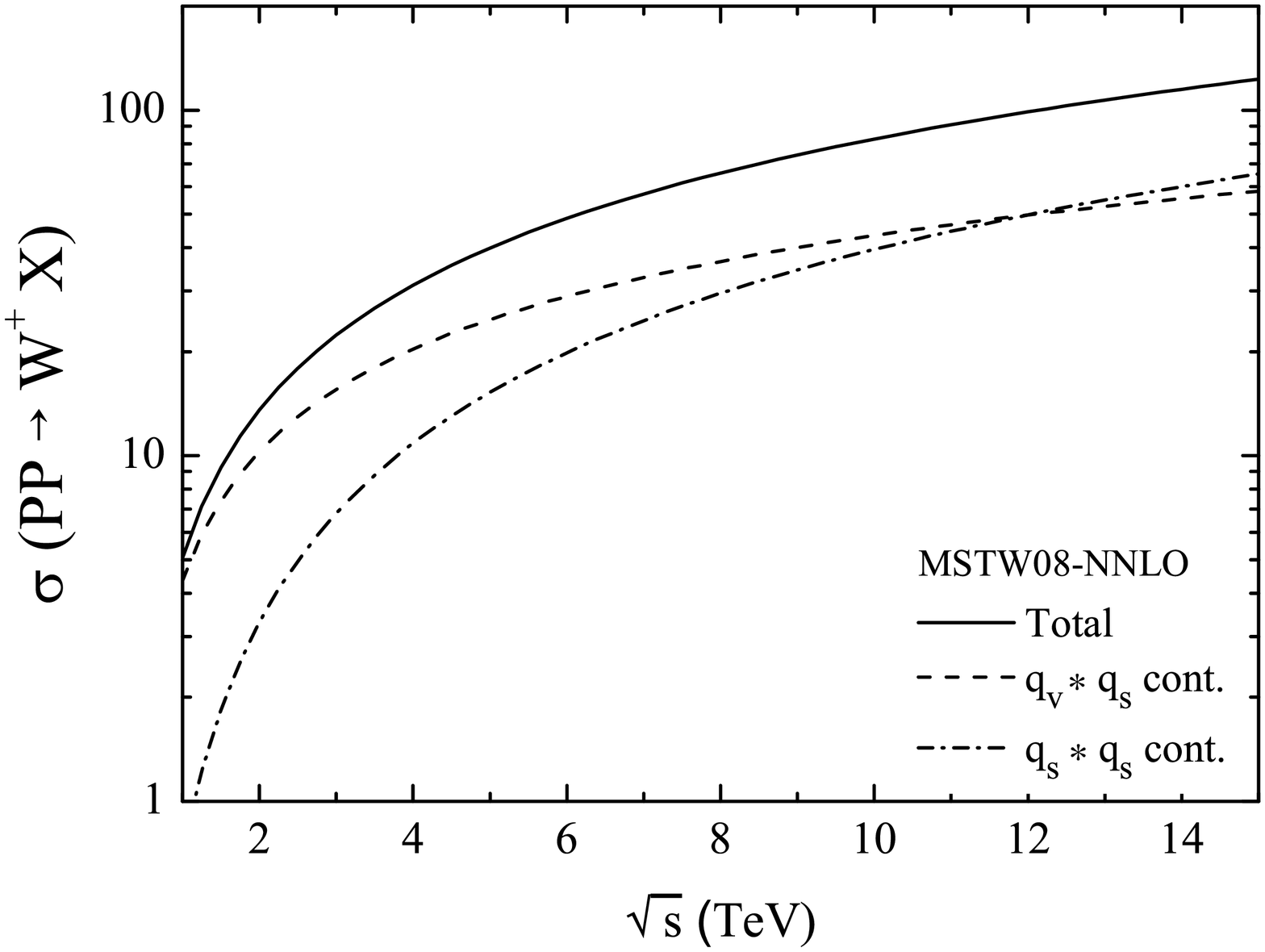}
\includegraphics[scale = 0.23]{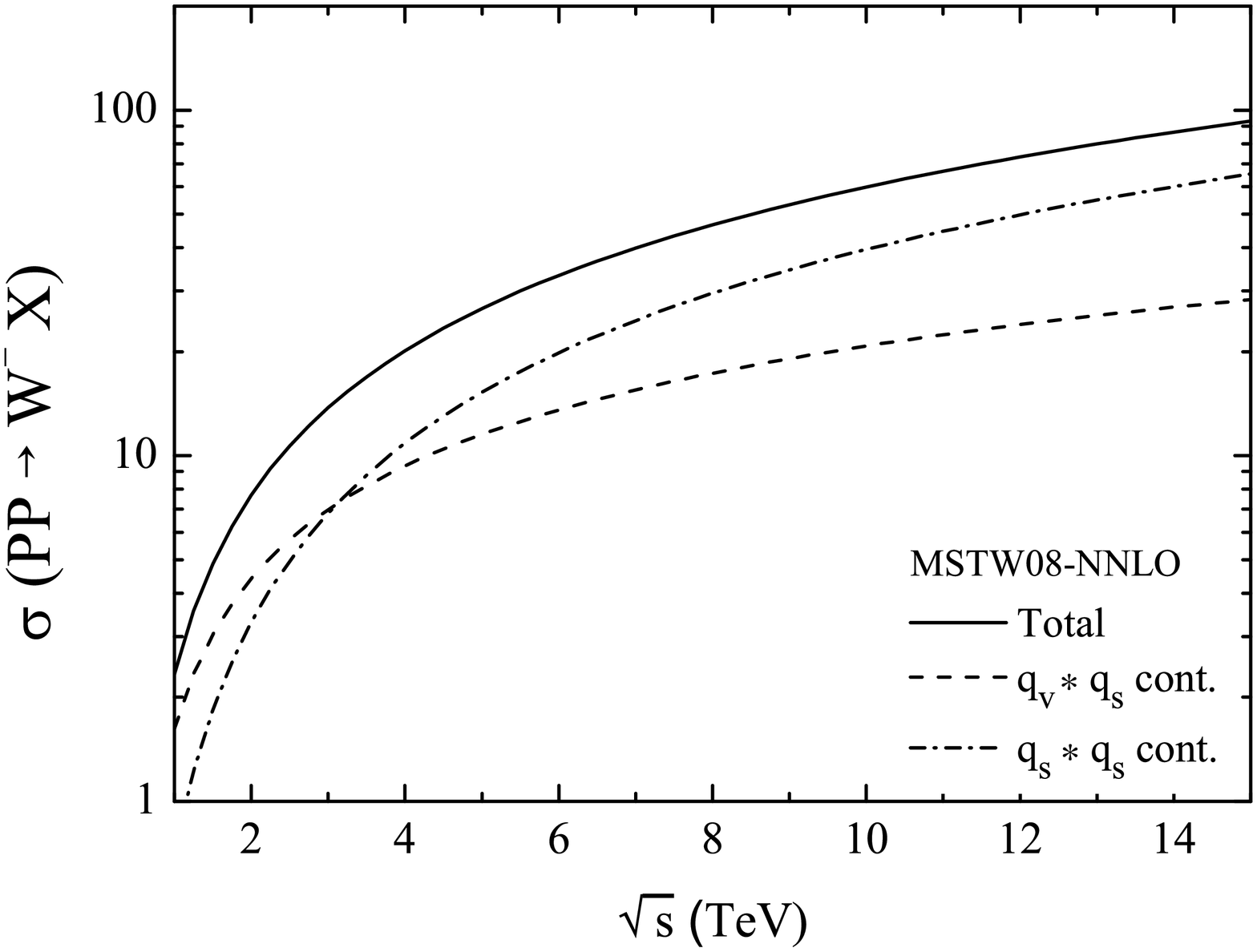}
\includegraphics[scale = 0.23]{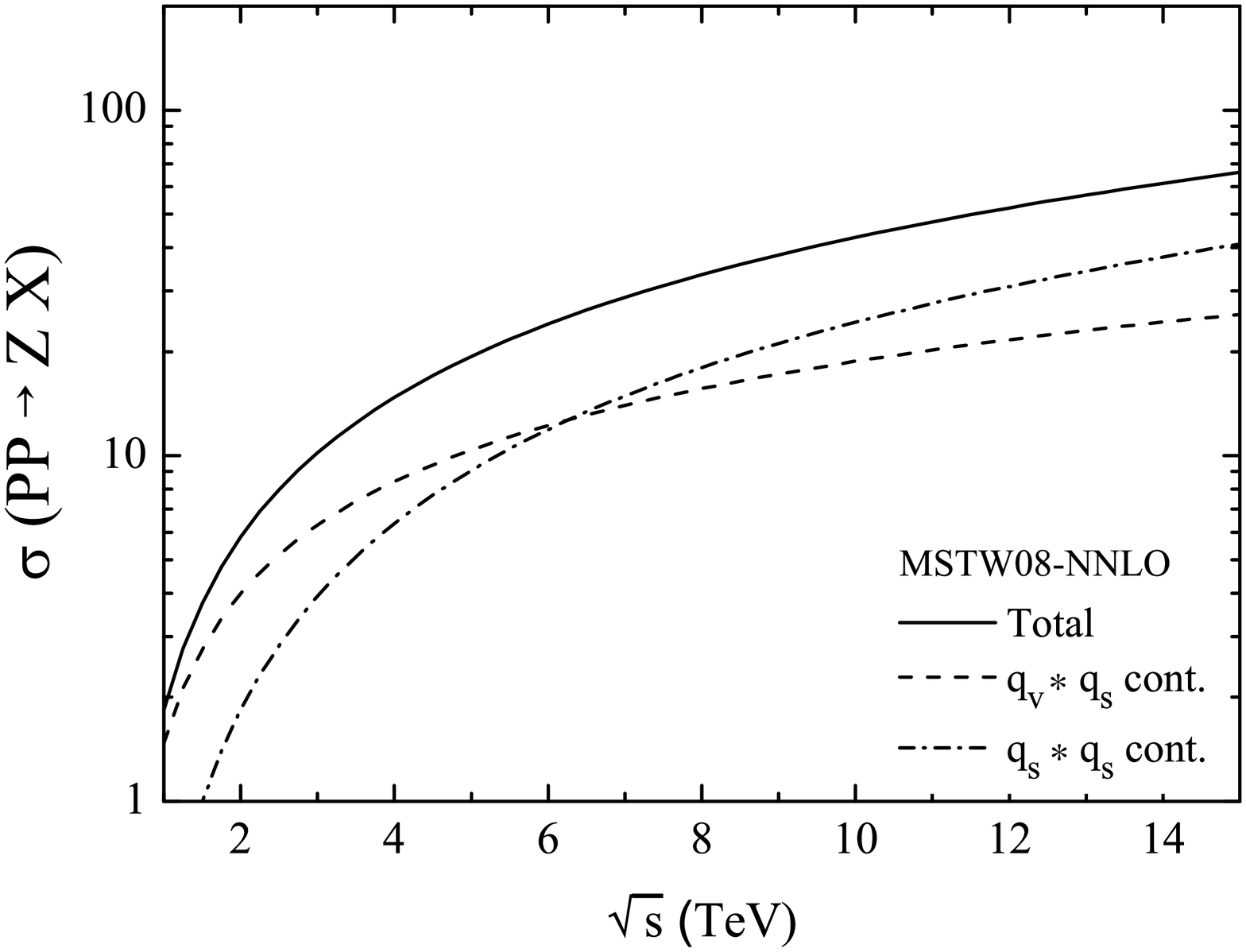}
\caption{The total cross sections for the production of weak bosons, $W^{+(-)}$ and $Z$
in proton-proton collision up to the full NNLO in QCD. The separate contributions from the valence-sea (dashed)
and the sea-sea (dot-dashed) interactions are also shown as function of energy.}%
\label{fig:sigma}%
\end{figure}
%
In Fig. \ref{fig:sigma}, we have calculated the cross sections for inclusive $W$ and $Z$ production up to the NNLO.
In evaluating the cross sections, we use the program in the VRAP \cite{VRAP} package,
which provides the rapidity distributions of W and Z bosons through the NNLO in QCD.
For the PDFs, we use the NNLO set of MSTW2008 PDFs.

The total cross section for $W^+$ production is different from that for $W^-$ production while they are identical in $p\bar{p}$ collision.
The separate contributions from the valence-sea and sea-sea quark interactions are also included.
There is no contribution from valence-valence quark interactions for $pp$ collisions up to the NLO,
and its contribution at the NNLO is small enough to be neglected.
Therefore, the role of the sea quarks is enhanced for $pp$ collisions and,
as energy increases, the sea quark contributions become increasingly important.
They are responsible for the increased production of $W^-$ and $Z$ relative to $W^+$.

In addition to the total cross section, we are here particularly interested
in the differential cross sections as a function of rapidity that sample the momentum
fractions of the partons and are therefore optimal for constraining the PDFs. 
Therefore, here we show  the differential cross section for $W$ and $Z$ boson production at leading order (LO),
{\it for the simplified schematic purpose}, in $pp$ collisions as,
\begin{eqnarray}
\nonumber
\frac{d \, \sigma}{dY}(pp \rightarrow W^+ X) &=&  \frac{2 \pi G_F}{3 \sqrt{2}} x_1 x_2 \times  \\
\nonumber
 && \{ \left|V_{ud}\right|^2 [u(x_1)\bar{d}(x_2) + \bar{d}(x_1)u(x_2)]
 + \left|V_{us}\right|^2 [u(x_1)\bar{s}(x_2) + \bar{s}(x_1)u(x_2)]\\
\nonumber
 && + \left|V_{cs}\right|^2 [c(x_1)\bar{s}(x_2) + \bar{s}(x_1)c(x_2)]
 + \left|V_{cd}\right|^2 [c(x_1)\bar{d}(x_2) + \bar{d}(x_1)c(x_2)]\\
\nonumber
 && + \left|V_{ub}\right|^2 [u(x_1)\bar{b}(x_2) + \bar{b}(x_1)u(x_2)]	
 + \left|V_{cb}\right|^2 [c(x_1)\bar{b}(x_2) + \bar{b}(x_1)c(x_2)] \} \ ,
\end{eqnarray}
\begin{eqnarray}
\label{eq:wdsdy}
\frac{d \, \sigma}{dY}(pp \rightarrow W^+ X) &=& \frac{2 \pi G_F}{3 \sqrt{2}} x_1 x_2 \times  \\
\nonumber
 && \{ \left|V_{ud}\right|^2 [\bar{u}(x_1)d(x_2) + d(x_1)\bar{u}(x_2)]
+ \left|V_{us}\right|^2 [\bar{u}(x_1)s(x_2) + s(x_1)\bar{u}(x_2)]\\
\nonumber
 && \left|V_{cs}\right|^2 [\bar{c}(x_1)s(x_2) + s(x_1)\bar{c}(x_2)]
+ \left|V_{cd}\right|^2 [\bar{c}(x_1)d(x_2) + d(x_1)\bar{c}(x_2)]\\
\nonumber
 && + \left|V_{ub}\right|^2 [\bar{u}(x_1)b(x_2) + b(x_1)\bar{u}(x_2)]
+ \left|V_{cb}\right|^2 [\bar{c}(x_1)b(x_2) + b(x_1)\bar{c}(x_2)] \}  \ ,
\end{eqnarray}
\begin{eqnarray}
\label{eq:zdsdy}
\nonumber
\frac{d \, \sigma}{dY}(pp \rightarrow Z X) &=&  \frac{2 \pi G_F}{3 \sqrt{2}} x_1 x_2 \times  \\
\nonumber
 && \{ g_u^2 [u(x_1)\bar{u}(x_2) + \bar{u}(x_1)u(x_2)
 + c(x_1)\bar{c}(x_2) + + \bar{c}(x_1)c(x_2)]\\
 && + g_d^2 [d(x_1)\bar{d}(x_2) + \bar{d}(x_1)d(x_2)
+  s(x_1)\bar{s}(x_2) +  \bar{s}(x_1)s(x_2) \\
\nonumber
 && + b(x_1)\bar{b}(x_2) +  \bar{b}(x_1)b(x_2)] \}  \ ,
\end{eqnarray}
where
\be
g_u^2 &=& (1-8\sin^2\theta_W/3 +32\sin^4\theta_W/9)/2 \ , \nonumber \\
g_d^2 &=& (1-4\sin^2\theta_W/3 +8\sin^4\theta_W/9)/2 \ .
\label{eq:gcpl}
\ee

Here $G_F$ is the Fermi coupling constant, and $|V_{ij}|$ are the Cabibbo-Kobayashi-Maskawa (CKM) matrix elements.
The parton momentum fractions, $x_1$ and $x_2$ are related to rapidity $Y$ by
\be
x_1 = \frac{M_{W,Z}}{\sqrt{s}}\ e^Y \ , \ \
x_2 = \frac{M_{W,Z}}{\sqrt{s}}\ e^{-Y} \ ,
\label{eq:x12}
\ee
where $M_{W,Z}$ are the masses of the $W$ and $Z$ bosons. Thus, $Y$ covers the range $-\rm{ln}(\sqrt{s}/M_{W,Z})$ to $\rm{ln}(\sqrt{s}/M_{W,Z})$.
For numerical evaluation, we use $M_W$ = 80.399 GeV, $M_Z$ = 91.188 GeV, $G_F$ = 1.166 $\times {10}^{-5} \, \rm{GeV}^{-2}$, and $\sin^2 \theta_W$ = 0.23 \cite{PDG10}.
As for the total cross sections, the factorization scale and the renormalization scale
are set to be the mass of the $W/Z$ boson, $\mu_F = \mu_R = M_{W/Z}$.

Please note that at the NNLO the initial gluon  as well as the heavy quark flavors  PDFs also contribute to $W$ and $Z$ production.
For the numerical evaluation, we have included gluons as well as all flavors up to b-quark.
For the PDFs, the CTEQ6.6 is the fit for NLO, while MSTW2008 provides the NNLO PDF set,
therefore, we compute the rapidity distributions of the W and Z boson differential cross sections, by using the VRAP program \cite{VRAP},
up to the NLO for CTEQ6.6 PDF set, and to the NNLO for MSTW2008 PDFs.

\begin{figure}[ht]%
\centering
\includegraphics[scale = 0.33]{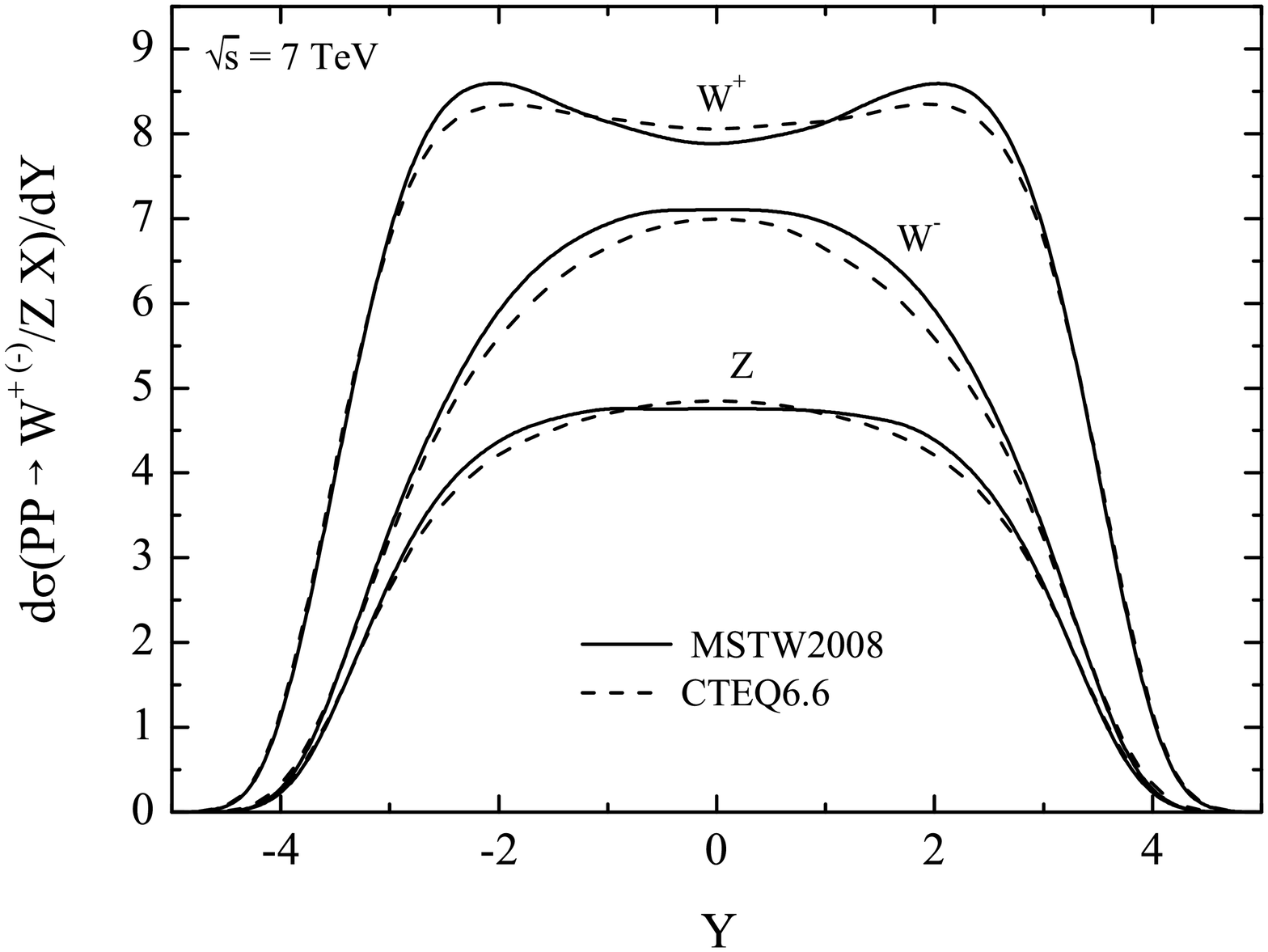}
\includegraphics[scale = 0.33]{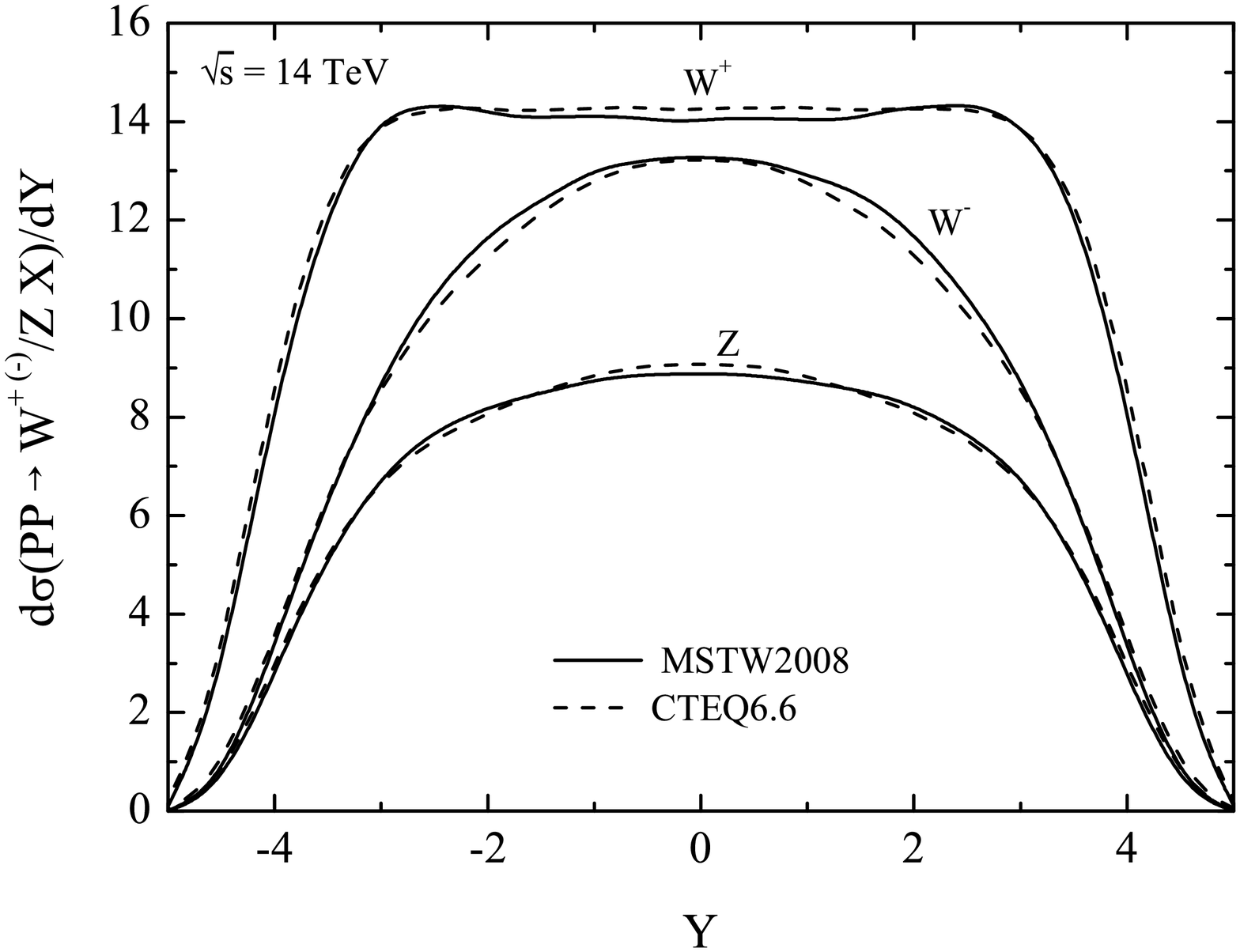}
\caption{The rapidity distribution of the differential cross sections for $W^{\pm}$ and Z
boson production. The left and right figures are for energies of $\sqrt{s}$ = 7 TeV
and $\sqrt{s}$ = 14 TeV, respectively. The results are presented for MSTW2008 (solid) \cite{MSTW08}
and CTEQ6.6 (dashed) \cite{CTEQ66} PDFs. }%
\label{fig:dsdy}%
\end{figure}

Fig. \ref{fig:dsdy} shows the distribution of the differential cross sections in rapidity
for both the CTEQ6.6 \cite{CTEQ66} and MSTW2008 \cite{MSTW08} PDFs, and for two different center-of-mass energy $\sqrt{s}$ = 7 and 14 TeV.
The differential cross sections in $pp$ collisions are symmetric with respect to the sign of the rapidity with $d \sigma_{W^\pm} (Y)$ = $d \sigma_{W^\pm} (-Y)$, while $d \sigma_{W^-} (Y)$ = $d \sigma_{W^+} (-Y)$ in $p\bar{p}$ collisions.
The accessible rapidity range is $|Y| \lsim 4.3$ at $\sqrt{s}$ = 7 TeV, and $|Y| \lsim 5$ at $\sqrt{s}$ = 14 TeV allowing for the exploration of parton momentum fractions $x$ as low as 1.7 $\times 10^{-4}$ and 4.2 $\times 10^{-5}$, respectively.
The range of the rapidity and momentum fractions is much broader than for the Tevatron whose rapidity range from $-3$ to 3 probed minimum values of $x_1$ and $x_2$ of 2 $\times 10^{-3}$ at $\sqrt{s}$ = 1.96 TeV.

Considering the LO cross section to be schematic, the $W/Z$ cross sections in $pp$ collisions
are from the valence-sea and the sea-sea interactions.
For the $pp$ process, the sea-sea interactions for $W^+$ and $W^-$ are the same,
thus the valence-sea interactions give rise to the difference
in the magnitude between $W^+$ and $W^-$ production.
And we note that because the CKM  elements squared have values of 0.95
for $|V_{ud}|^2$ and $|V_{cs}|^2$, and 0.05 for $|V_{us}|^2$ and $|V_{cd}|^2$,
the valence-sea contributions are dominated by the $u_v \bar{d}$ channel for $W^+$
and $\bar{u} d_v$ for $W^-$ production.
The distribution function of $u_v$, $u_v (x)$, is higher than $d_v(x)$
while $\bar{u}$ and $\bar{d}$ are almost the same,
therefore the cross section of $W^+$ production is higher than that of $W^-$ production.

In addition, the sea-sea contributions are dominant in the central rapidity region,
while the valence-sea contributions dominate in the forward and backward regions.
This cause the difference in the shape of $W^\pm$ distributions.
As the energy increases the accessible $x$ values become smaller.
Consequently, the sea-sea contributions are more important
than valence-sea contributions at the high energy,
and this effect dominates at the central rapidity region.
One can see this in Fig. \ref{fig:dsdy}, especially with $W^\pm$ rapidity distributions.
Considering the results with the MSTW2008 PDFs for example,
at $\sqrt{s}$ = 7 TeV, the $W^+$ differential cross section has two peaks near $|Y|$ = 2.
However, these peaks are disappeared due to the enhancement of sea-sea contribution
at $\sqrt{s}$ = 14 TeV.
Also, the differential cross section of $W^+$ and $W^-$ are closer at higher energy,
especially at $Y$ = 0.


Finally we note that the $b$ quark interactions have an effect on the $Z$ distributions
as large as 4.1 \% and 5.8 \% in the central region, and  contribute less than 1 \% at $|Y| \gsim 2.5$
and $|Y| \gsim 3.3$ for $\sqrt{s}$ = 7 TeV and 14 TeV, respectively.
On the other hand, for $W$ production, the $b$ quark contributions are much more suppressed
because of the small CKM matrix elements squared,
with values of $10^{-5}$ for $|V_{ub}|^2$ and $2 \times 10^{-3}$ for $|V_{cb}|^2$.
Therefore, we can totally ignore the $b$ quark contributions
for the charge asymmetry of $W^\pm$ production, and safely ignore it for the ratio of $W/Z$.


\section{The Charge Asymmetry of $W^\pm$ Production}

\begin{figure}[ht]%
\centering
\includegraphics[scale = 0.35]{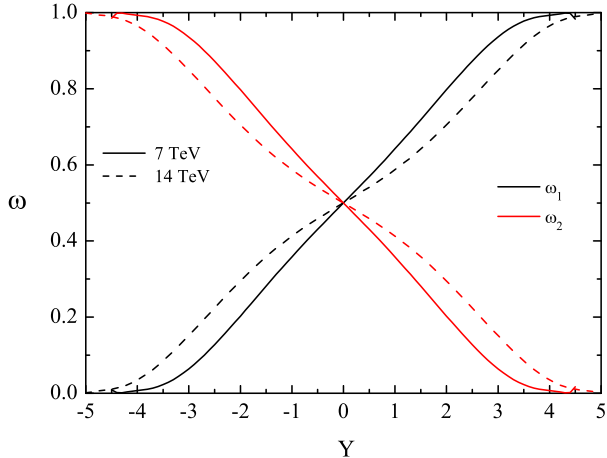}
\caption{The weights $\omega_1$ (dark black) and $\omega_2$ (bright red) as a function of rapidity for $\sqrt{s}$ = 7 and 14 TeV.}%
\label{fig:wgts}%
\end{figure}

In this section, we investigate the rapidity dependence of the charge asymmetry.
The $W^+$ is produced mainly by the $u$ and $\bar{d}$, and $W^-$ by $\bar{u}$ and $d$ channels.
Since the PDFs of these quarks are different, a charge asymmetry appears in the $W^+$ and $W^-$ production distributions, which is defined as
\be
A_W(Y)&=& \frac{d \sigma_{W^+}/d Y - d \sigma_{W^-}/d Y}{d \sigma_{W^+}/d Y + d \sigma_{W^-}/d Y} \ .
\label{eq:Ay}
\ee
The charge asymmetry in $pp$ collision can be approximated in terms of the $u$ and $d$ quark distributions as
\be
A(Y) &\approx& \biggl\{\omega_1 \frac{u(x_1)-d(x_1)}{u(x_1)+d(x_1)}
 + \omega_2 \frac{u(x_2)-d(x_2)}{u(x_2)+d(x_2)} \biggr\} \ ,
\label{eq:ay-simp1}
\ee
where the weight $\omega_1$ and $\omega_2$ are
\be
\nonumber
\omega_1 &\equiv& \frac{\frac{\bar{u}(x_2)+\bar{d}(x_2)}{u(x_2)+d(x_2)}}  {\left[\frac{\bar{u}(x_1)+\bar{d}(x_1)}{u(x_1)+d(x_1)} + \frac{\bar{u}(x_2)+\bar{d}(x_2)}{u(x_2)+d(x_2)} \right]}   \ , \\
\omega_2 &\equiv& \frac{\frac{\bar{u}(x_1)+\bar{d}(x_1)}{u(x_1)+d(x_1)}} {\left[\frac{\bar{u}(x_1)+\bar{d}(x_1)}{u(x_1)+d(x_1)} + \frac{\bar{u}(x_2)+\bar{d}(x_2)}{u(x_2)+d(x_2)} \right]}   \ .
\label{eq:ay-weight}
\ee
In Eq. \ref{eq:ay-simp1} we made the approximation that $\bar{u}(x)$ = $\bar{d}(x)$.
As noted before, the $b$ quark interactions contribute a negligible amount as a result of the very small values of the CKM matrix elements\footnote{We can perfectly neglect the $c$ quark contribution to the numerator of $A(Y)$;
however, in the denominator there are terms from $c \bar s$ and $\bar c s$
which cannot be ignored; they change the normalization.
We use the original definitions, Eqs. (6) and (14), for our numerical presentation.
}.
The weights $\omega_1$ and $\omega_2$ are shown in Fig. \ref{fig:wgts} as a function of $W$ rapidity.
Note that $\omega_1$ and $\omega_2$ both approach the value of 0.5
when $Y = 0$, i.e. $x_1$ = $x_2$. Also, $\omega_1$ approaches 1 (0)
for large positive (negative) rapidity, while $\omega_2$ approaches 1 (0)
for large negative (positive) rapidity.

The expression corresponding to Eq. \ref{eq:ay-simp1} for $p\bar{p}$ collisions is given by \cite{BHKW}
\be
\label{eq:ay-ppb}
A(Y) &\sim&
D^{-1}\left[\frac{u(x_1)-d(x_1)}{u(x_1)+d(x_1)} - \frac{u(x_2)-d(x_2)}{u(x_2)+d(x_2)}\right] \ .
\ee
The multiplication factor has a value of 0.91 in the central region \cite{BHKW}, in contrast with a value of 0.5 for the $\omega$ factors in $pp$ collisions.
{}From Eq. \ref{eq:ay-ppb}, it follows that $A(Y) = -A(-Y)$, while $A(Y) = A(-Y)$ for $pp$ collisions.
In summary, in $p\bar{p}$ collisions the $W^+$ distribution is symmetric with the $W^-$ distribution, $d\sigma_{W^\pm}(Y) = d\sigma_{W^\mp}(-Y)$.
Therefore,  the rate of the two is the same at $Y=0$.
In contrast, in $pp$ collisions, the rapidity distributions of $W^+$ and $W^-$ are each symmetric with respect to rapidity, $d\sigma_{W^\pm}(Y) = d\sigma_{W^\pm}(-Y)$, and their magnitudes do not coincide at zero rapidity.
Therefore, in the central region for $pp$ collision
\be
A(Y=0) &\approx&
\frac{u(x)-d(x)}{u(x)+d(x)} = \frac{u_v(x)-d_v(x)}{u_v(x)+d_v(x)+2S(x)}\ ,
\label{eq:ay}
\ee
with $S(x)$ = $\bar{u}(x)$ = $\bar{d}(x)$, where $x=M_W/\sqrt{s}$,  while $A(Y=0)$=0 in $p\bar{p}$ collisions.
This difference makes the determination of $u/d$ much useful by using $pp$ data.

Inclusively produced weak bosons decay to leptons, $W \rightarrow l \nu_l$ and $Z \rightarrow l \bar{l}$, and the rapidity $Y_{W,Z}$ are reconstructed from the energy ($E_l$) and the longitudinal momentum ($p_{l,L}$) of the secondary leptons.
The rapidity of the $Z$ boson can be fully reconstructed from the measurement of its decay products.
In contrast, the $W$ rapidity cannot be reconstructed because the longitudinal momentum of the neutrino is not measured.
%
We therefore reconstruct the $W$ rapidity from the momentum of the charged lepton and the (indirectly determined) transverse momentum of neutrino \cite{HHK}. We introduce the definitions
\be
Y_{\pm} = \frac{1}{2} {\rm ln}
\frac{E_l + p_{l,L} + p_{\nu_l,T}e^{Y_{\nu_\pm}}}
{E_l - p_{l,L} + p_{\nu_l,T}e^{-Y_{\nu_\pm}}},
\label{eq:ypm}
\ee
where
\be
Y_{\nu_\pm} = Y_l \pm {\rm ln}[1+\delta+\sqrt{\delta(2+\delta)}],
\label{eq:ynupm}
\ee
with
\be
\delta = \frac{M_W^2 - M^2_{T}}{2p_{l,T}p_{\nu_l,T}} \ .
\label{eq:del}
\ee
Here $M^2_{T}$ is the transverse mass of the lepton and neutrino defined as $M^2_{T} = (|p_{l,T}|+|p_{\nu,T}|)^2 - (p_{l,T}+p_{\nu,T})^2$ .

\begin{figure}[h]%
\centering
\includegraphics[scale = 0.35]{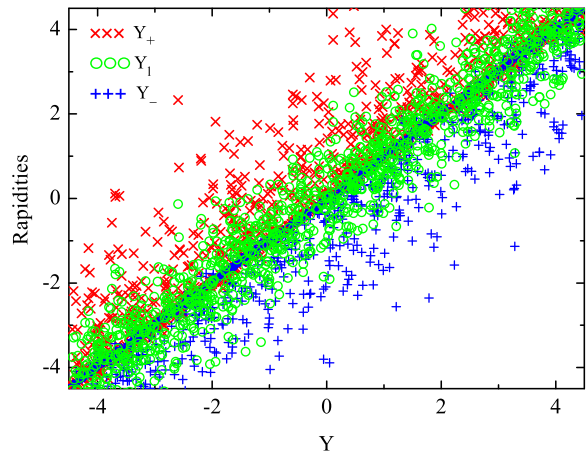}
\caption{The distributions of the reconstructed rapidities $Y_\pm$ and the charged lepton
rapidity $Y_l$ as function of $Y$ of $W$ boson.}%
\label{fig:Rap}%
\end{figure}

Fig. \ref{fig:Rap} shows the distribution of the reconstructed rapidity $Y_\pm$ and the charged lepton rapidity $Y_l$ as a function of the rapidity of $W$.
$Y_l$ and $Y_\pm$ are related: $Y_l = (Y_+ + Y_-)/2$.
In the massless limit, the lepton rapidity $Y_l$ is the same as the pseudorapidity $\eta$, which is used to investigate the lepton charge asymmetry \cite{CFG, BAMO, MP, CDF98ay, BFLMY, CDF05ay, D008aymu, D008aye, ATLAS11ay, CMS11ay, CMS12ay}.
As indicated in Fig. \ref{fig:Rap}, $Y_l$ correlates roughly with the rapidity $Y$,
therefore, $Y_l$ could also be an approximate optional choice of $Y$ for the numerical purpose,
even though $Y_l$ cannot be used in the analytic expressions.
In this paper, however, we use the experimentally fully reconstructible $Y_\pm$ to investigate the charge asymmetry and the ratio of $W/Z$.
As can be seen from Eq. (11), we can choose either $Y_+$ or $Y_-$ as the appropriate observable
to study the charge asymmetry.
In this paper, we simply take $Y_-$ and will present results.


\begin{figure}[h]%
\centering
\includegraphics[scale = 0.3]{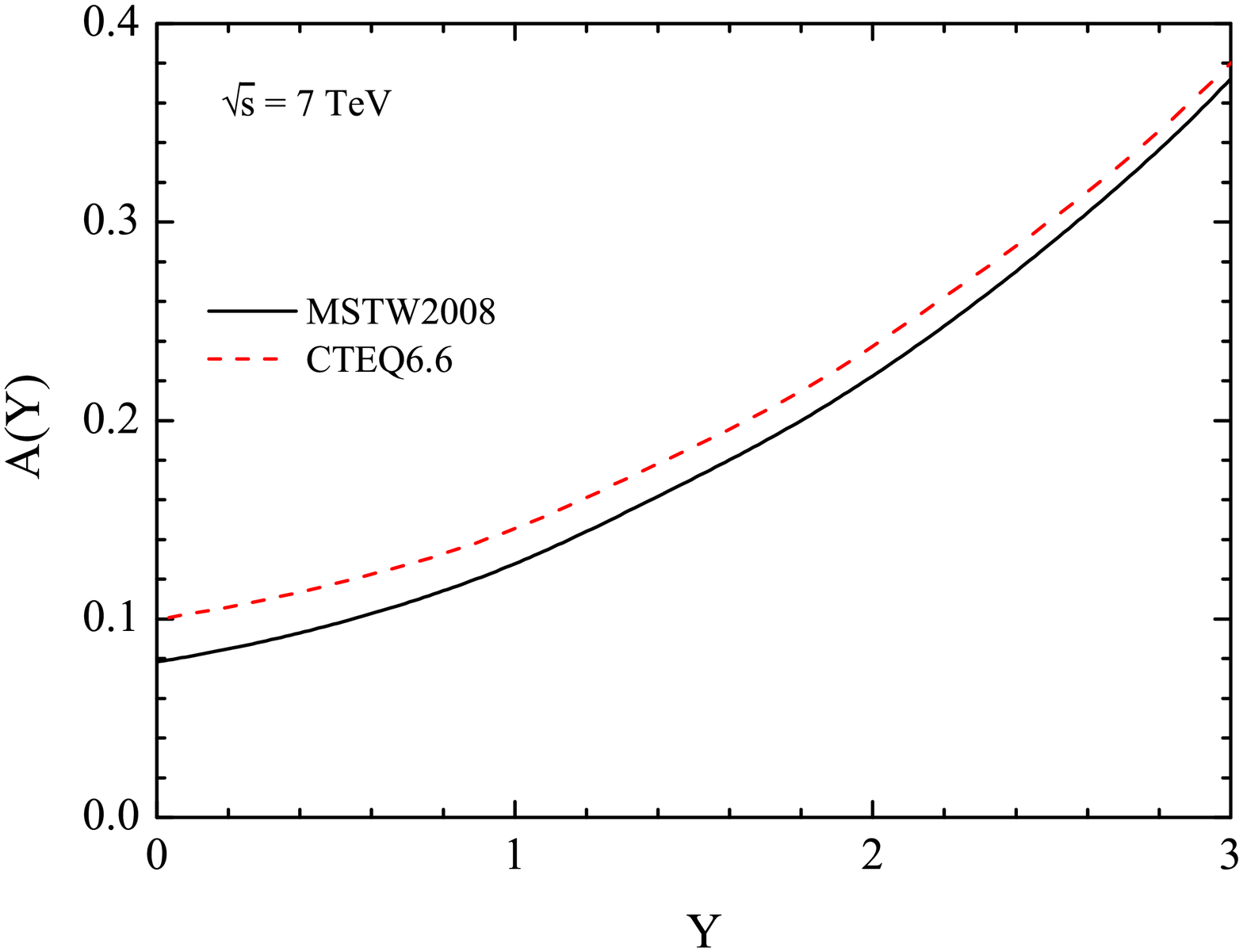}
\includegraphics[scale = 0.3]{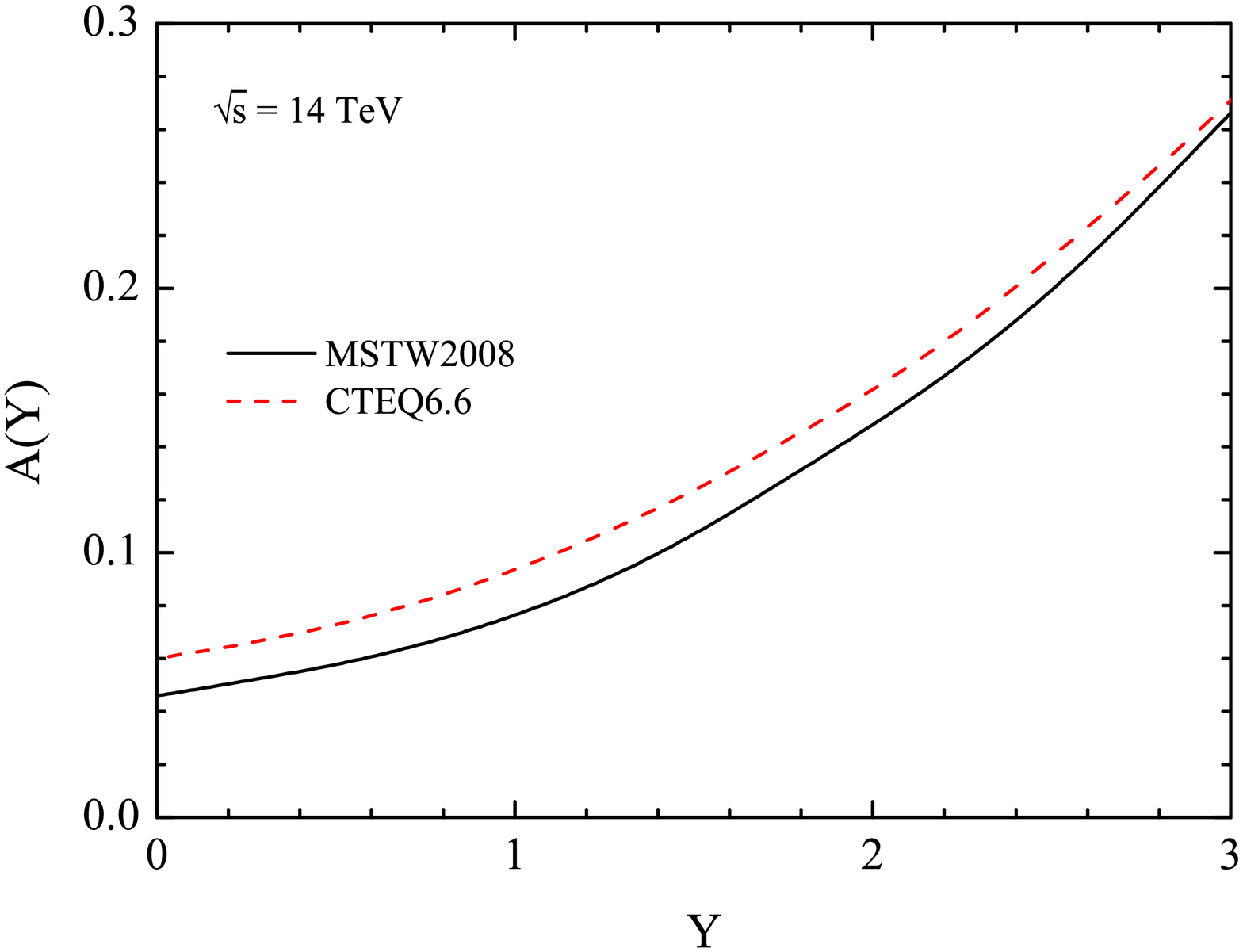}
\caption{ The charge asymmetry $A(Y)$ for the different PDFs for energies of $\sqrt{s}$ = 7 and 14 TeV
as a function of  the reconstructed rapidity $Y_{-}$.
The results for the PDFs are differentiated as follows:
black (solid) for MSTW2008 \cite{MSTW08}, and red (dashed) for CTEQ6.6 \cite{CTEQ66}.}%
\label{fig:Aypm}%
\end{figure}
We now redefine the $W$ charge asymmetry $A(Y)$ in terms of the reconstructed rapidity $Y_-$ as
\be
A(Y)&=& \frac{d\sigma_{W^+}/dY_{-} - d\sigma_{W^-}/dY_{-}}
{d\sigma_{W^+}dY_{-} + d\sigma_{W^-}/dY_{-}} \ .
\label{eq:Aypm}
\ee
The results for the asymmetry following Eq. (\ref{eq:Aypm}) are presented in Fig. \ref{fig:Aypm}
for different sets of PDFs.
As for the rapidity distributions in Sec. II, these results are fully calculated
up to NLO for CTEQ6.6 and NNLO for MSTW2008 PDF sets, including gluon contributions.
As shown in the figure, the results of the $W$ charge asymmetry with CTEQ6.6 and MSTW2008 PDFs
have a substantial difference: the difference at the central rapidity is 22 \% at $\sqrt{s}$ = 7 TeV
and 23 \% at $\sqrt{s}$ = 14 TeV.
This difference is reduced to 3 \% and 1 \% at $Y$ = 3
for the energy of $\sqrt{s}$ = 7 TeV and 14 TeV, respectively.

We also note that the charge asymmetry for $\sqrt{s}$ = 14 TeV is smaller than
the results for $\sqrt{s}$ = 7 TeV at the same rapidity.
At higher energies, the accessible momentum fraction, $x$, becomes lower.
At the low $x$ region, the values of $u$ and $d$ quark distributions,
are increased while the difference between  $u(x)$ and $d(x)$ are reduced.
Therefore, the W charge asymmetry decreases as the collision energy increases.

In addition, although we do not present in the figure, we evaluated the LO charge asymmetry
with the consistent PDF set of MSTW2008 \cite{MSTW08}, namely MSTW2008-LO PDFs.
At $\sqrt{s}$ = 7 TeV, {\it the difference between the NNLO and LO charge asymmetry} is about 12 \%
at the central rapidity, and it decreases with the rapidity up to $Y \sim$ 2.
At the large rapidities, $Y \gsim$ 2, the NNLO results have difference
0 \% - 4 \% from the LO results.
On the other hand, at $\sqrt{s}$ = 14 TeV, the NNLO charge asymmetry to the LO result
has about 3 \% difference at the zero rapidity,
and their differences for whole rapidities range from 0 \% to 5 \%.

\section{Charm Quark Distributions of the Proton}
Next, we will focus on the charm quark distributions in the proton.
At LO, the charm quark contributions are from the interactions of
$c\bar{s}$, $c\bar{d}$ and $c\bar{b}$ ($\bar{c}s$, $\bar{c}d$ and $\bar{c}b$) for $W^+$($W^-$)
and of $c\bar{c}$ for $Z$.
As already mentioned in Sec. II, for $W$ production the terms
proportional to $|V_{cd}|^2$ and $|V_{cb}|^2$ are quite small,
therefore the dominant subprocess is $c \bar{s}$ ($s \bar{c}$) for $W^+$ ($W^-$).

\begin{figure}[ht]%
\centering
\includegraphics[scale = 0.23]{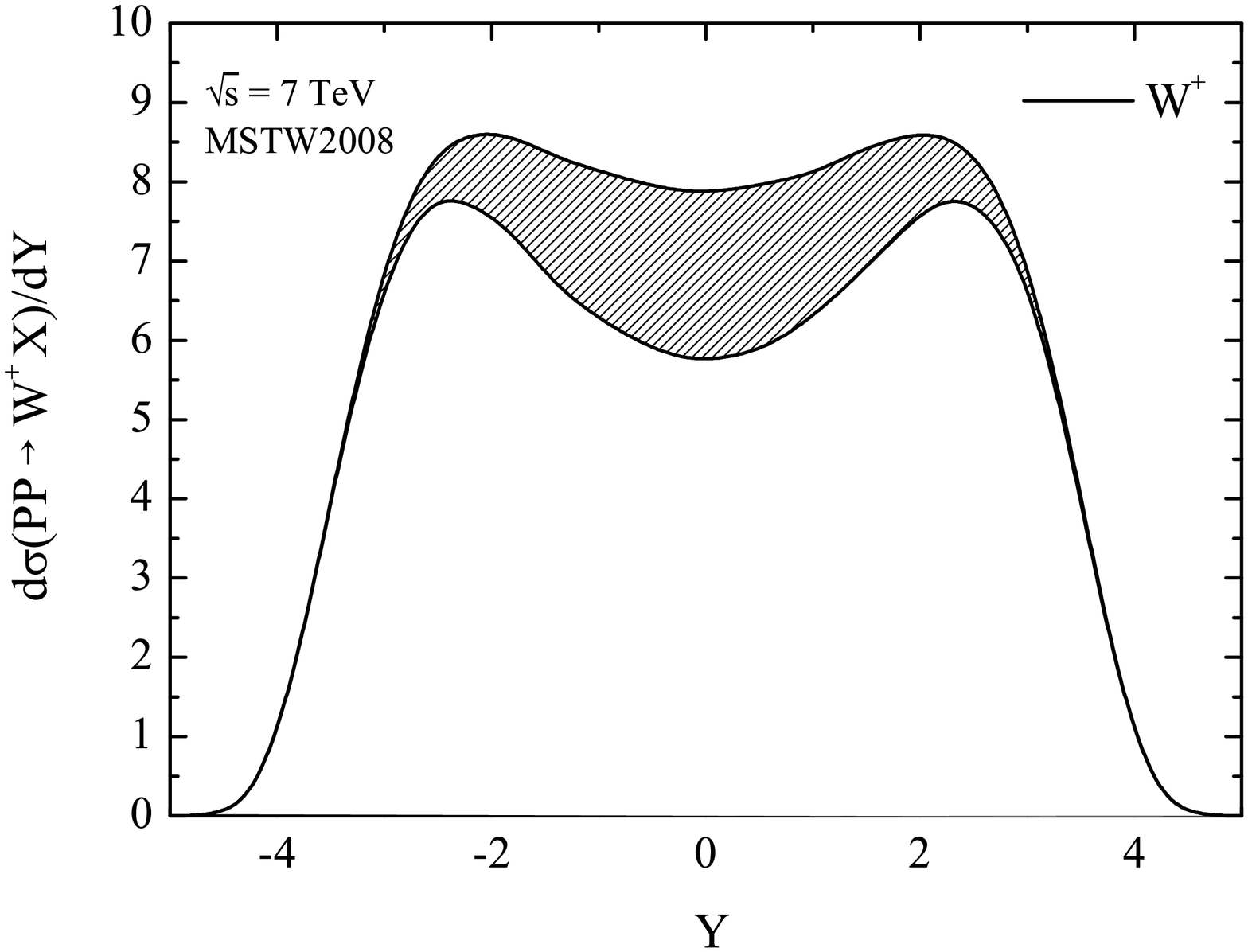}
\includegraphics[scale = 0.23]{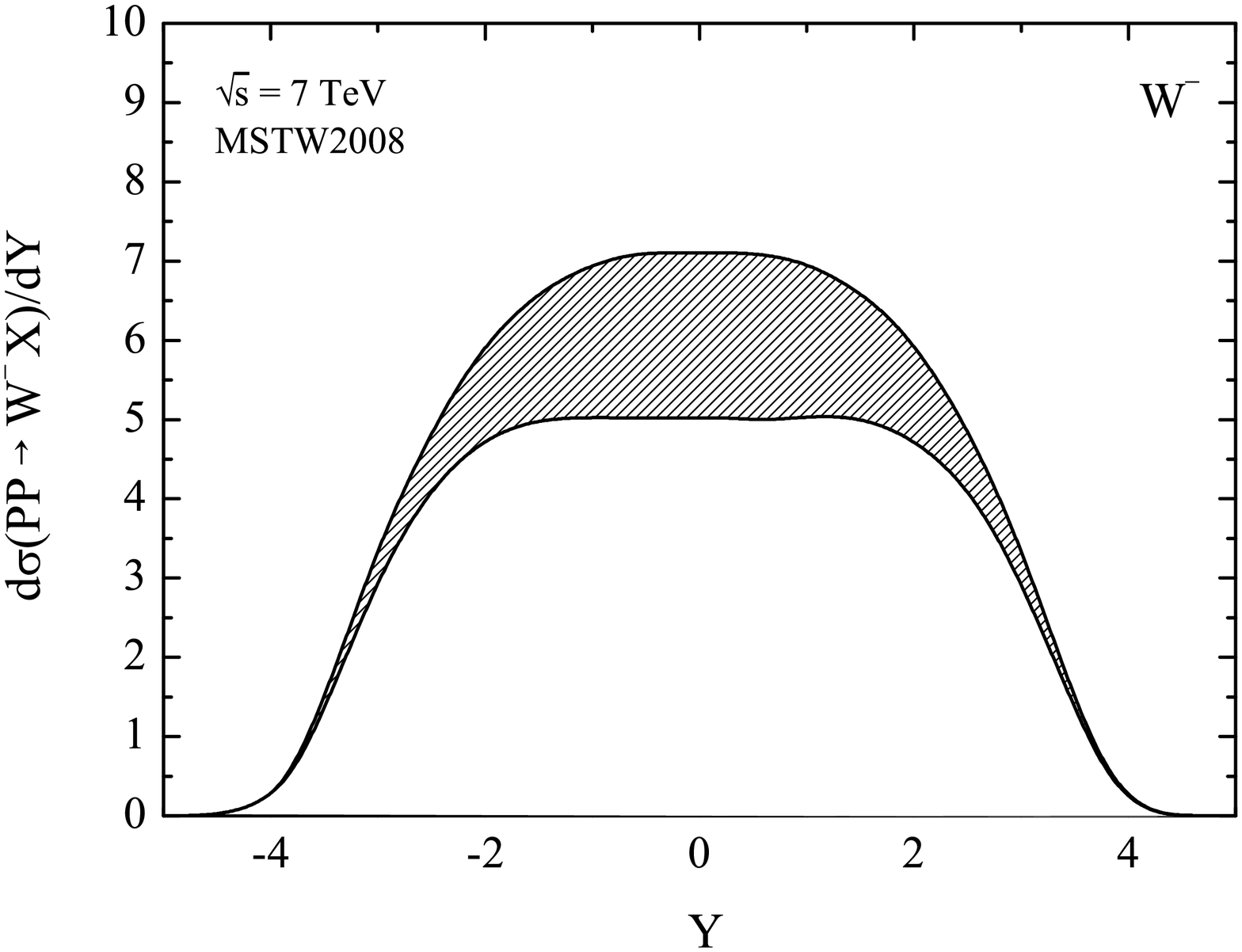}
\includegraphics[scale = 0.23]{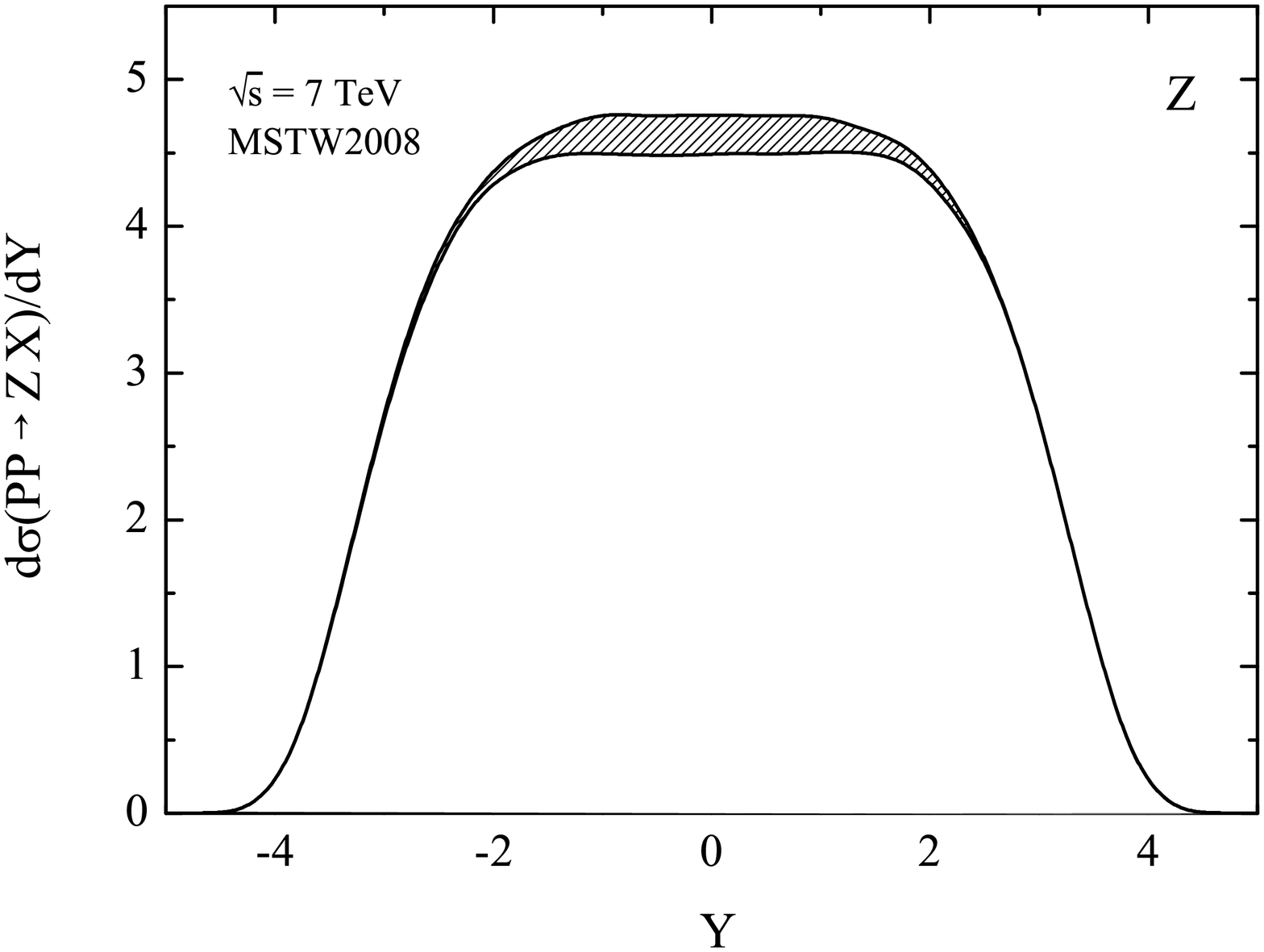}
\includegraphics[scale = 0.23]{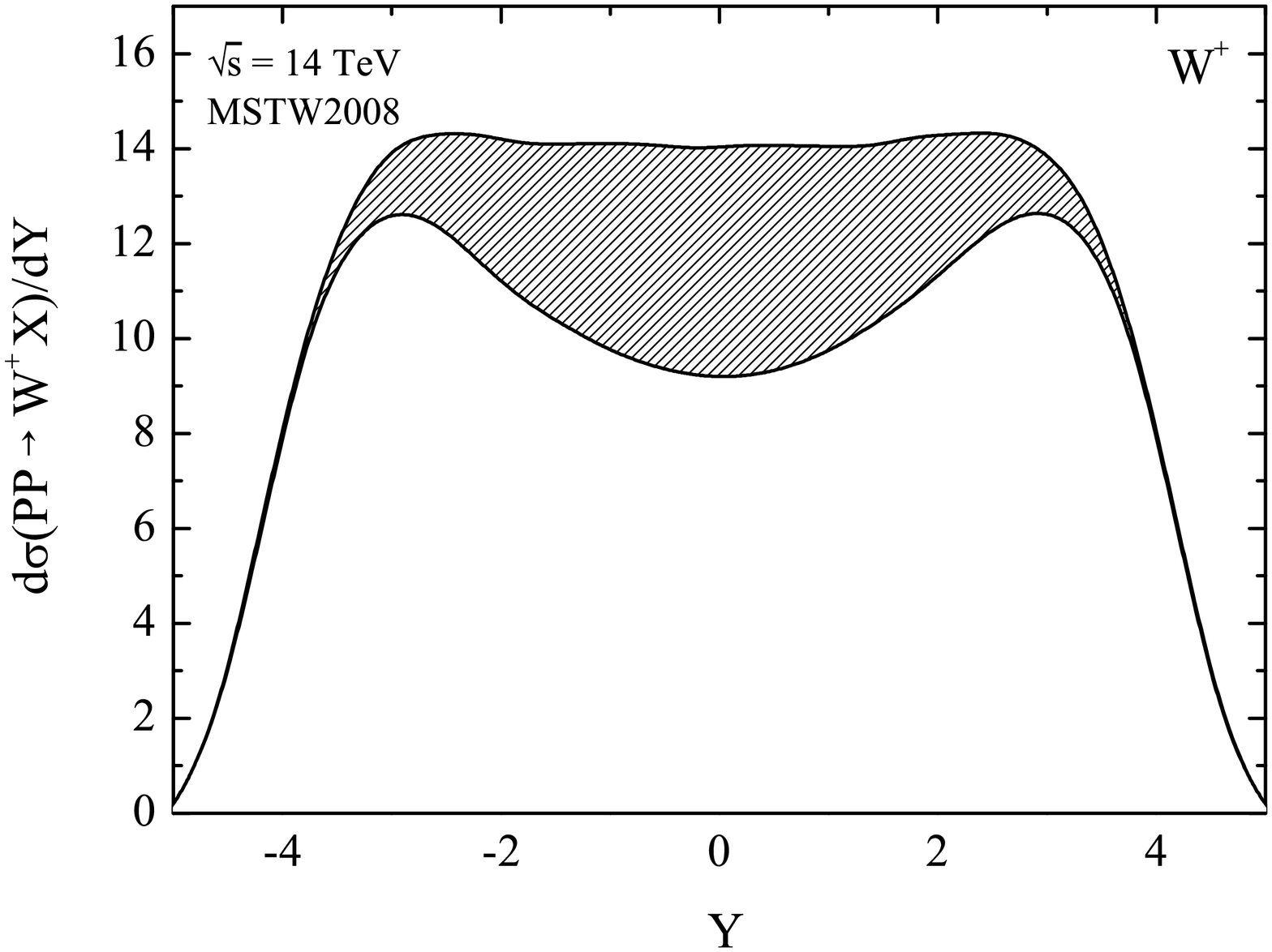}
\includegraphics[scale = 0.23]{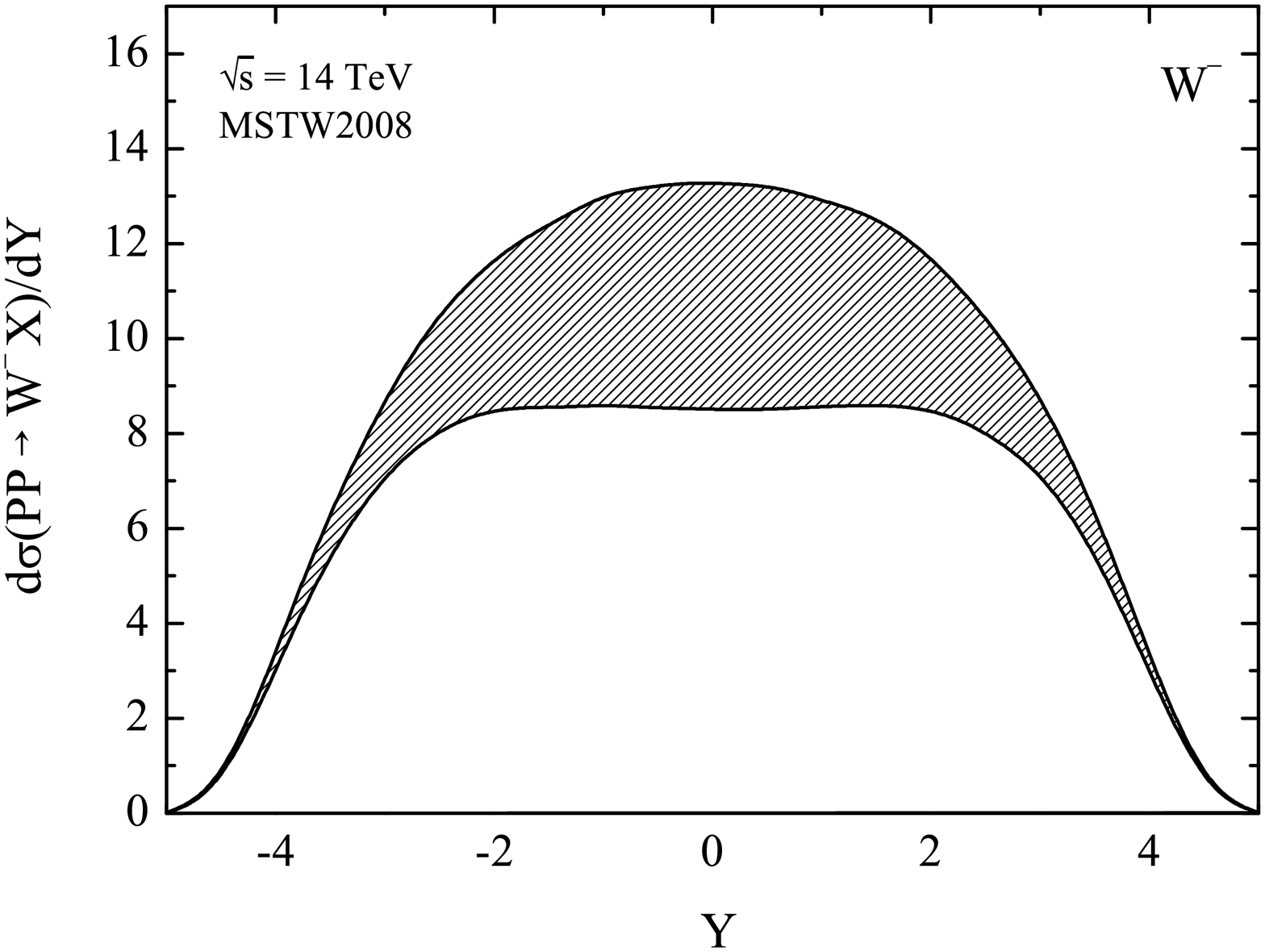}
\includegraphics[scale = 0.23]{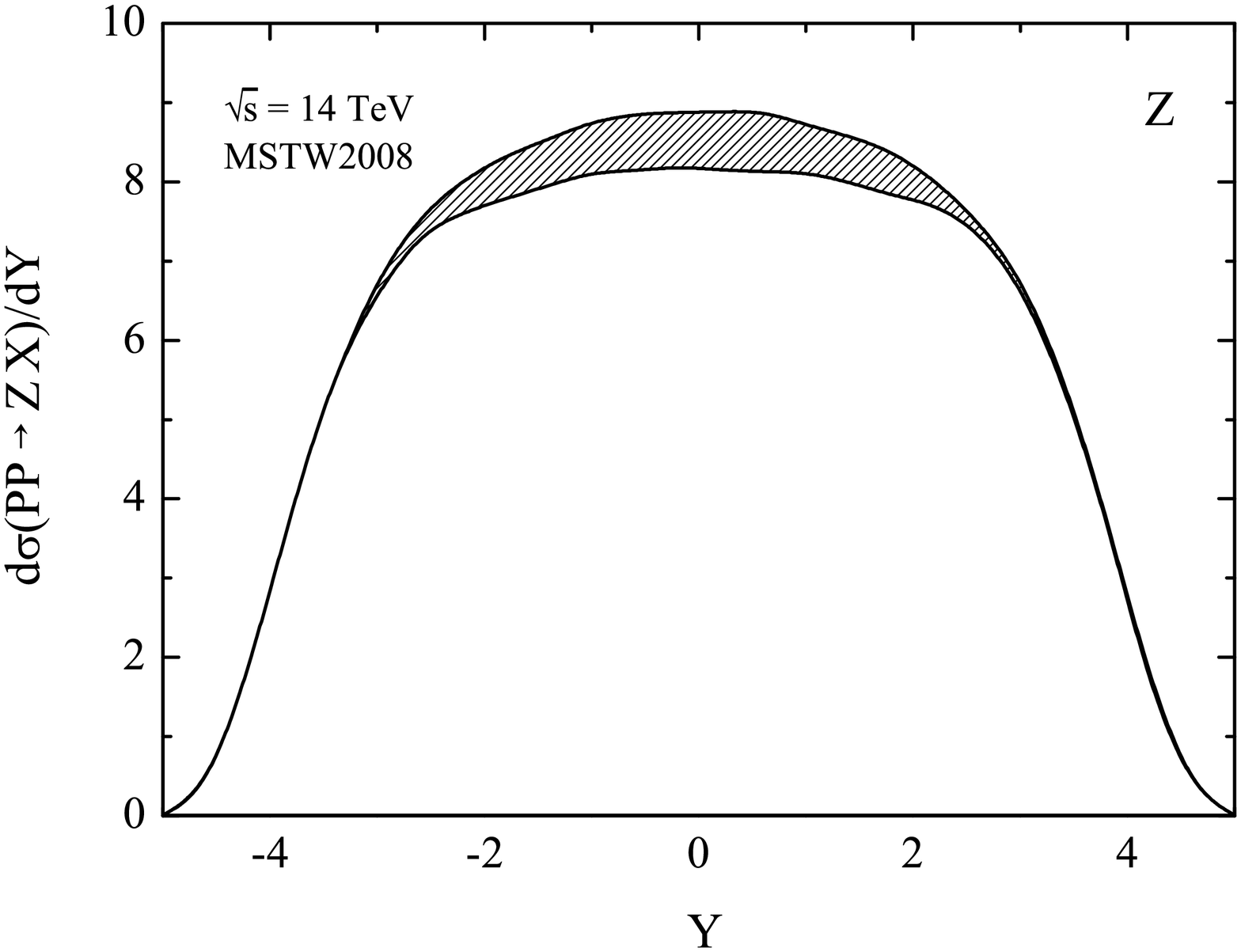}
\caption{The charm quark contribution to the rapidity distribution of $d \sigma / dY$
for the production of $W^+$, $W^-$, and $Z$ boson with $\sqrt{s}$ = 7 TeV and 14 TeV
at the full NNLO in QCD.
}%
\label{fig:dsdy-c}%
\end{figure}

In Fig. \ref{fig:dsdy-c}, we present the rapidity distributions of $W^\pm$ and $Z$ production
with and without the charm PDF, evaluated completely at the full NNLO using the VRAP program.
Thus, the shaded area displays the charm quark contributions
to the rapidity distributions of the total weak boson cross section.
This charm contribution dominates in the central region and is quite large as much as 27\% to $W^+$,
29\% to $W^-$, and 6\% to $Z$ production cross sections at $\sqrt{s}$ = 7 TeV.
At 14 TeV its influence increases to 34\%, 36\%, and 8\%, respectively.
Therefore, it is important to constrain the charm quark distribution in the proton more precisely.

The charm structure function has been examined with the charm production cross section data in deep inelastic scattering (DIS) processes at the HERA \cite{H1, Boroun}.
Ref. \cite{Boroun} suggested the ratio of $F^{c\bar{c}}_L$ and $F^{c\bar{c}}_2$
to extract the charm structure functions from the reduced cross section.
In this paper, we will investigate the charm PDF using the ratio
of the $W$ and $Z$ differential rapidity distributions.

First, we define the quantity $B(Y)$, introduced in Ref. \cite{HHK} for $p\bar{p}$ collisions,
in terms of the experimentally reconstructed rapidity $Y_-$ as
\be
B(Y) = \frac{d\sigma_{W^+}/dY_-+ d\sigma_{W^-}/dY_-}
{d \sigma_Z/dY} \ .
\label{eq:By}%
\ee
We found that unlike the charge asymmetry $A(Y)$, the newly defined ratio $B(Y)$ is very insensitive for orders in QCD.
Numerical differences between the results at the LO, NLO and NNLO are only less than 1 \%
for both energies, $\sqrt{s}$ = 7 and 14 TeV.
Therefore, please note that the $B(Y)$ can be analyzed directly using the LO expression,
Eqs. (2-3), although our numerical results in this paper are all evaluated at the full NNLO.
The almost identical $B(Y)$ for orders implies
that there is little uncertainty from the QCD corrections as well as
all other uncertainties such as from the PDFs of light quark flavors, the factorization scale,
detector systematics and etc.
Thus, the quantity  $B(Y)$ is expected to reveal quite precisely
the charm distribution at the LHC.

\begin{figure}[h]%
\centering
\includegraphics[scale = 0.35]{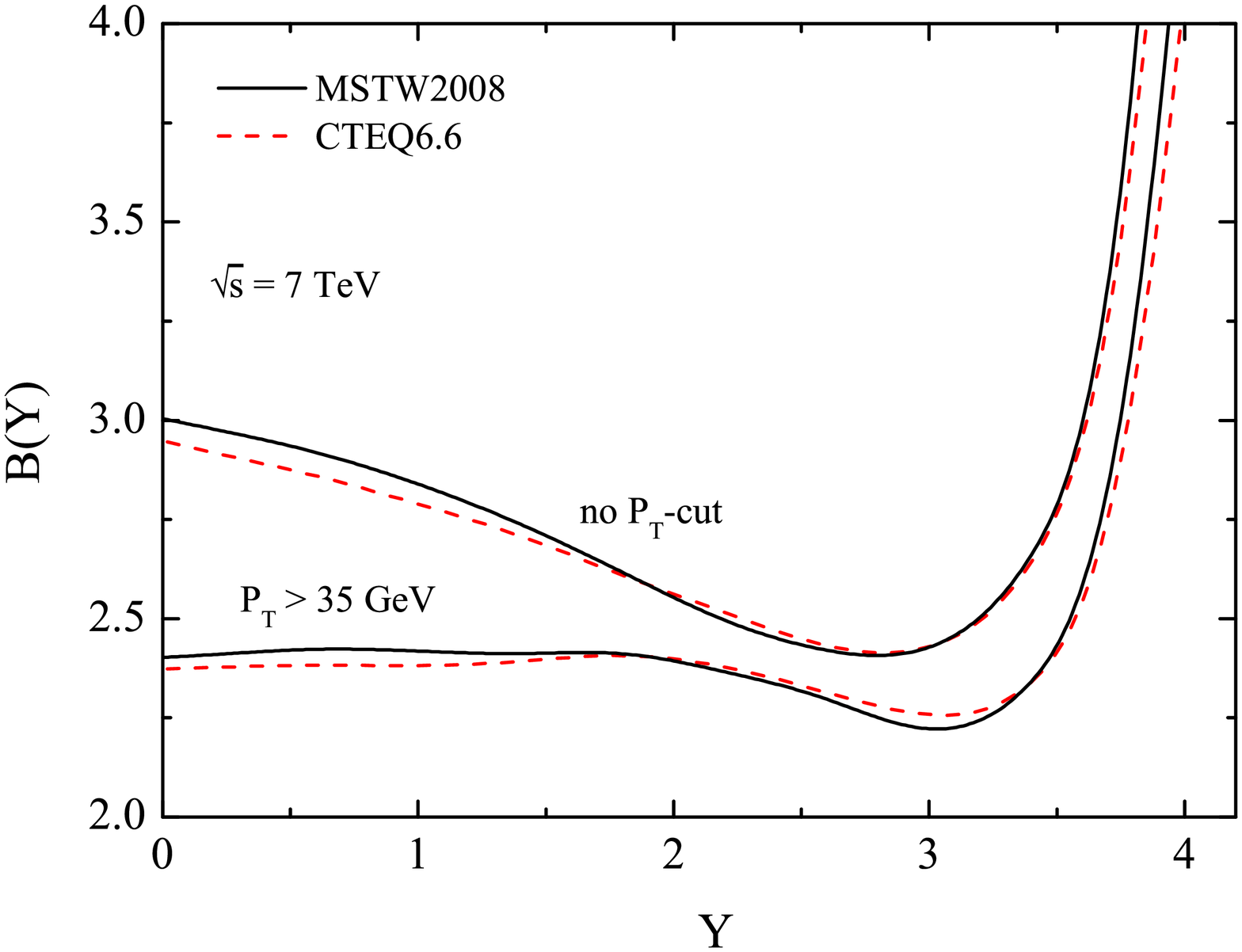}
\includegraphics[scale = 0.35]{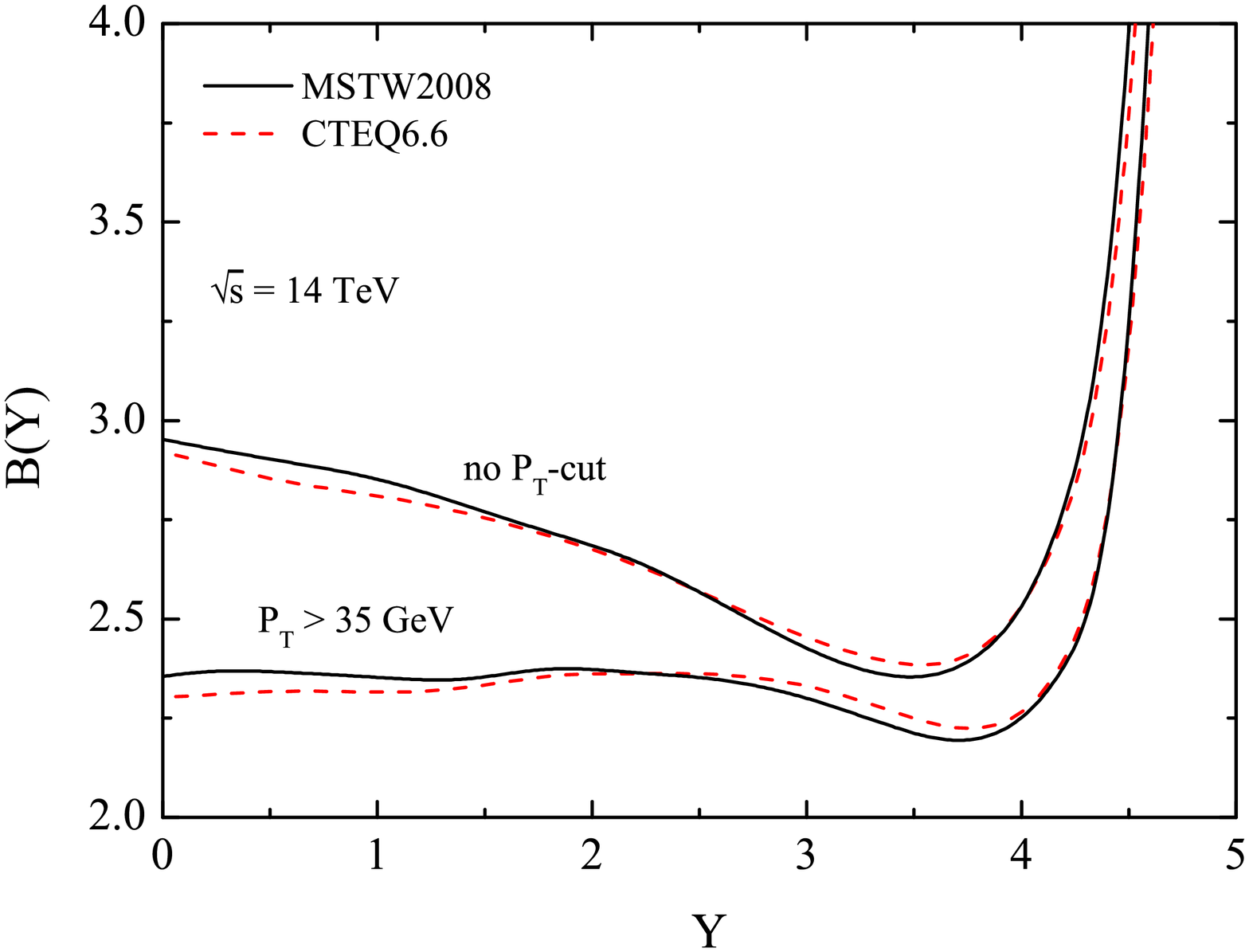}
\caption{$B(Y)$ as a function of rapidity for $\sqrt{s}$ = 7 and 14 TeV.
The distributions without $P_T$ cut are displayed with the results evaluated with $P_T >$ 35 GeV.
As in Fig. \ref{fig:Aypm}, the MSTW2008 \cite{MSTW08} results are shown in black solid curves
and CTEQ6.6 PDF \cite{CTEQ66} as red-dashed lines. }%
\label{fig:b-ypm}%
\end{figure}

In Fig. \ref{fig:b-ypm} we show $B(Y)$ for the different PDFs, CTEQ6.6 and MSTW2008,
using the charm distribution in each PDF set.
For the CTEQ6.6 PDFs, the B(Y) has the value of $\sim$ 2.95 at $|Y|$ = 0
for both $\sqrt{s}$ = 7 and 14 TeV.
The B(Y) quantity decreases as the rapidity increases,
and from a certain value of high rapidity, it goes to the infinity
due to the small production of $Z$ boson.
For the MSTW2008 PDF set, the values of B(Y) at low rapidity are slightly higher
than the results for the CTEQ6.6 PDFs,
but its behavior for whole rapidity region is the same as that for CTEQ6.6 PDFs.
The effect of the cut on the transverse momentum, $P_T$, is also included.
Here, we select $P_T >$ 35 GeV as introduced
in the recent CMS measurement of charge asymmetry \cite{CMS12ay}.
The results with this $P_T$-cut applied are reduced approximately to 2.4 for $|Y| < 2$ at $\sqrt{s}$ = 7 TeV, and  2.3 - 2.4 for $|Y| < 3$ at $\sqrt{s}$ = 14 TeV for both CTEQ and MSTW PDFs.

\begin{figure}[ht]%
\centering
\includegraphics[scale = 0.35]{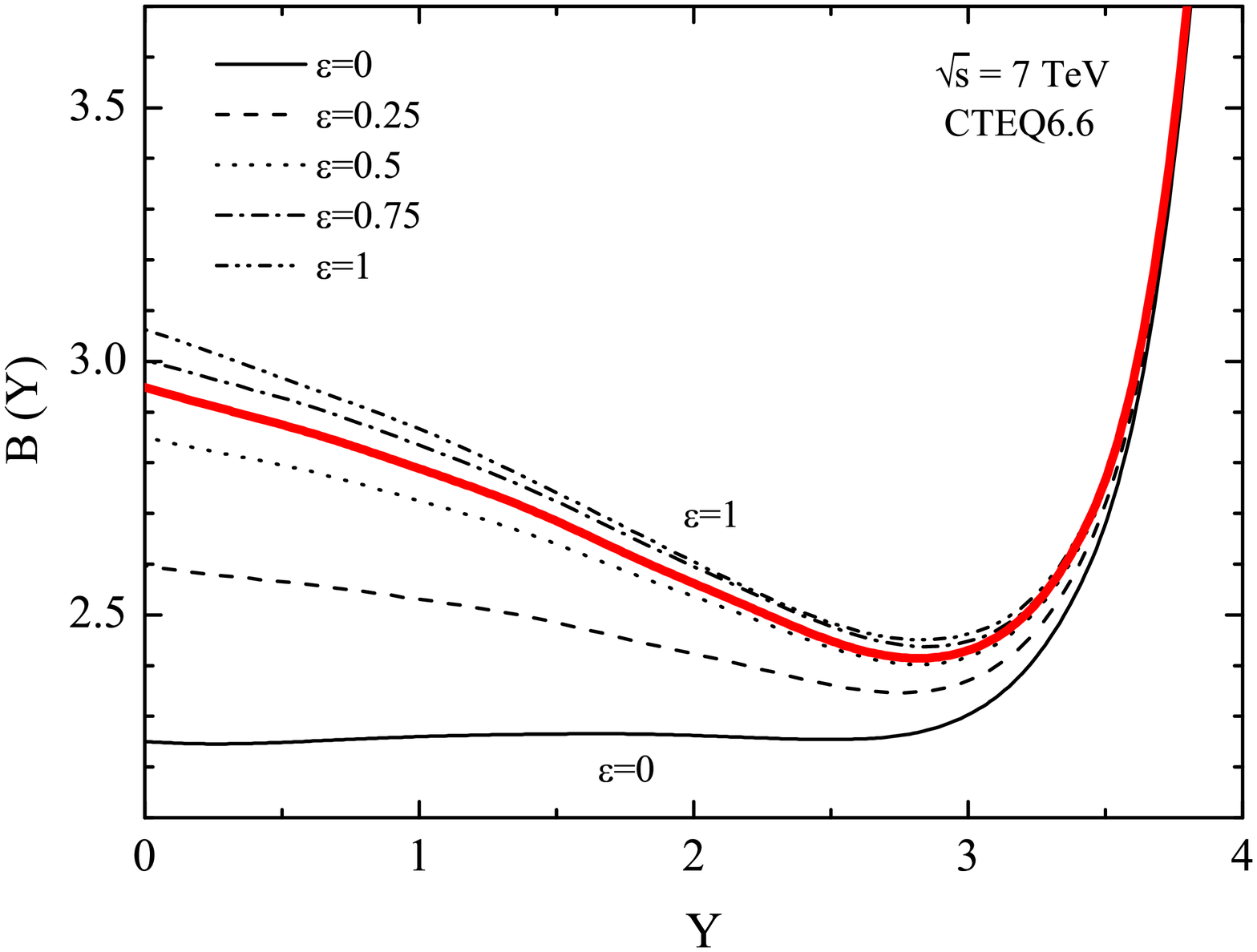}
\includegraphics[scale = 0.35]{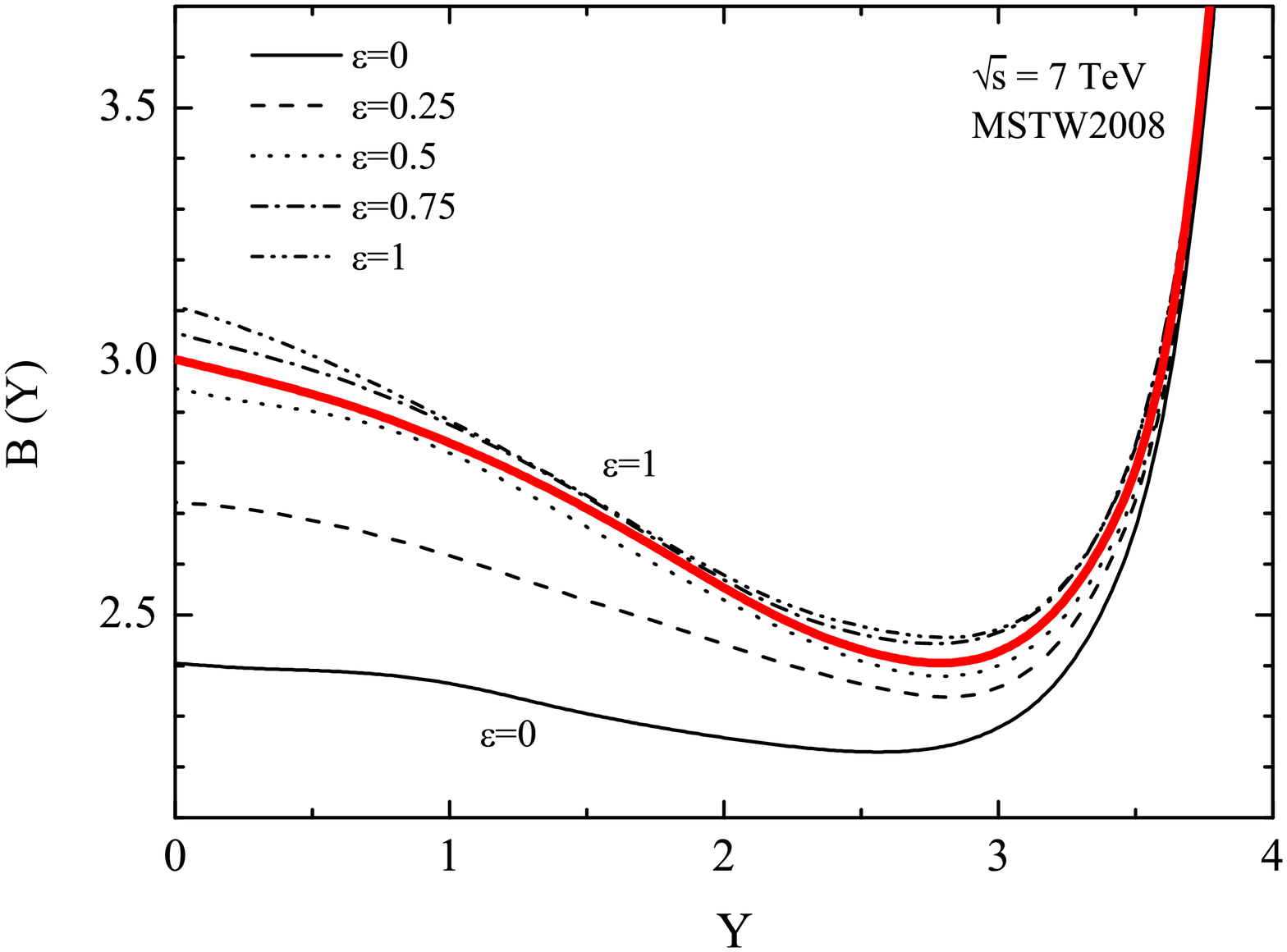}
\includegraphics[scale = 0.35]{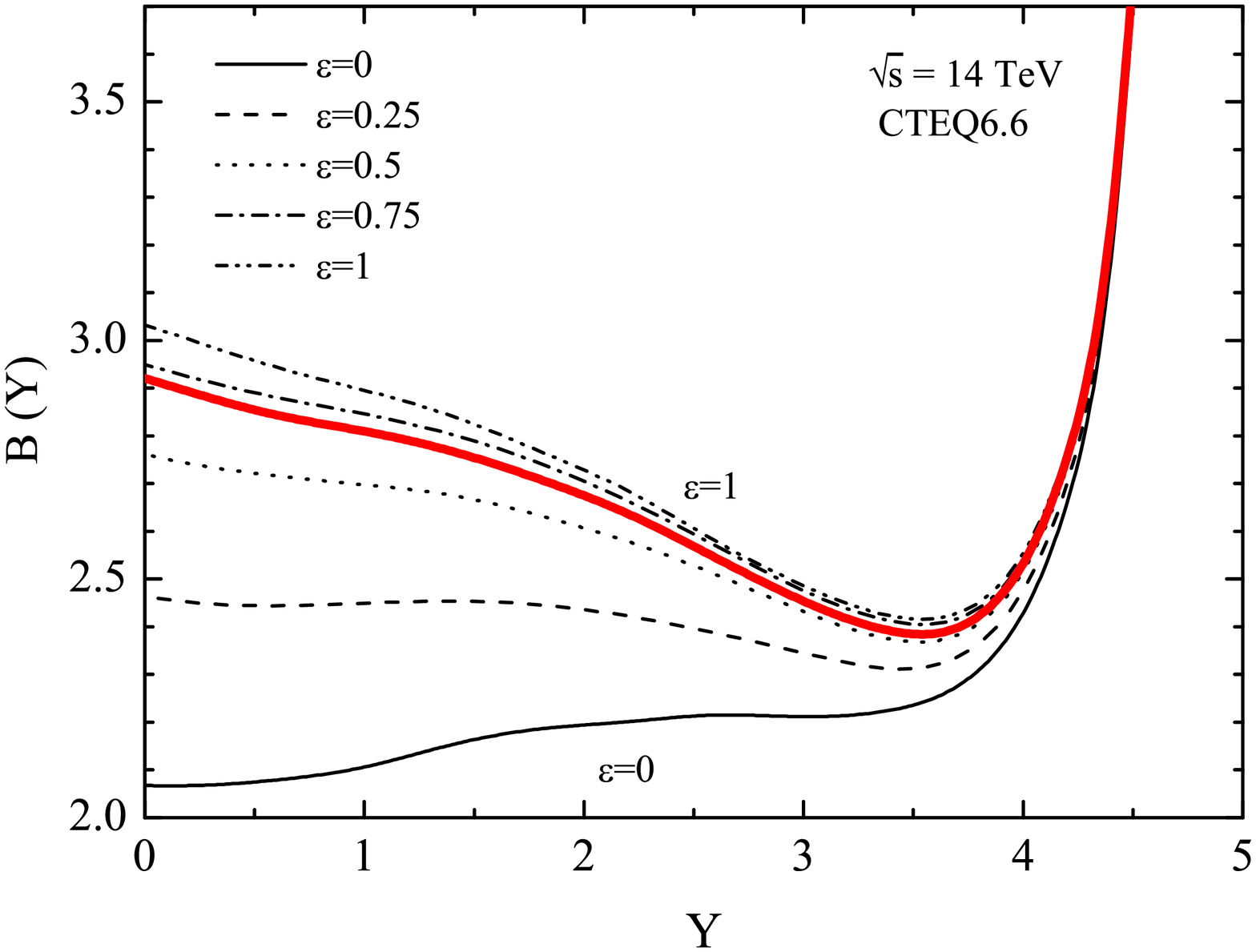}
\includegraphics[scale = 0.35]{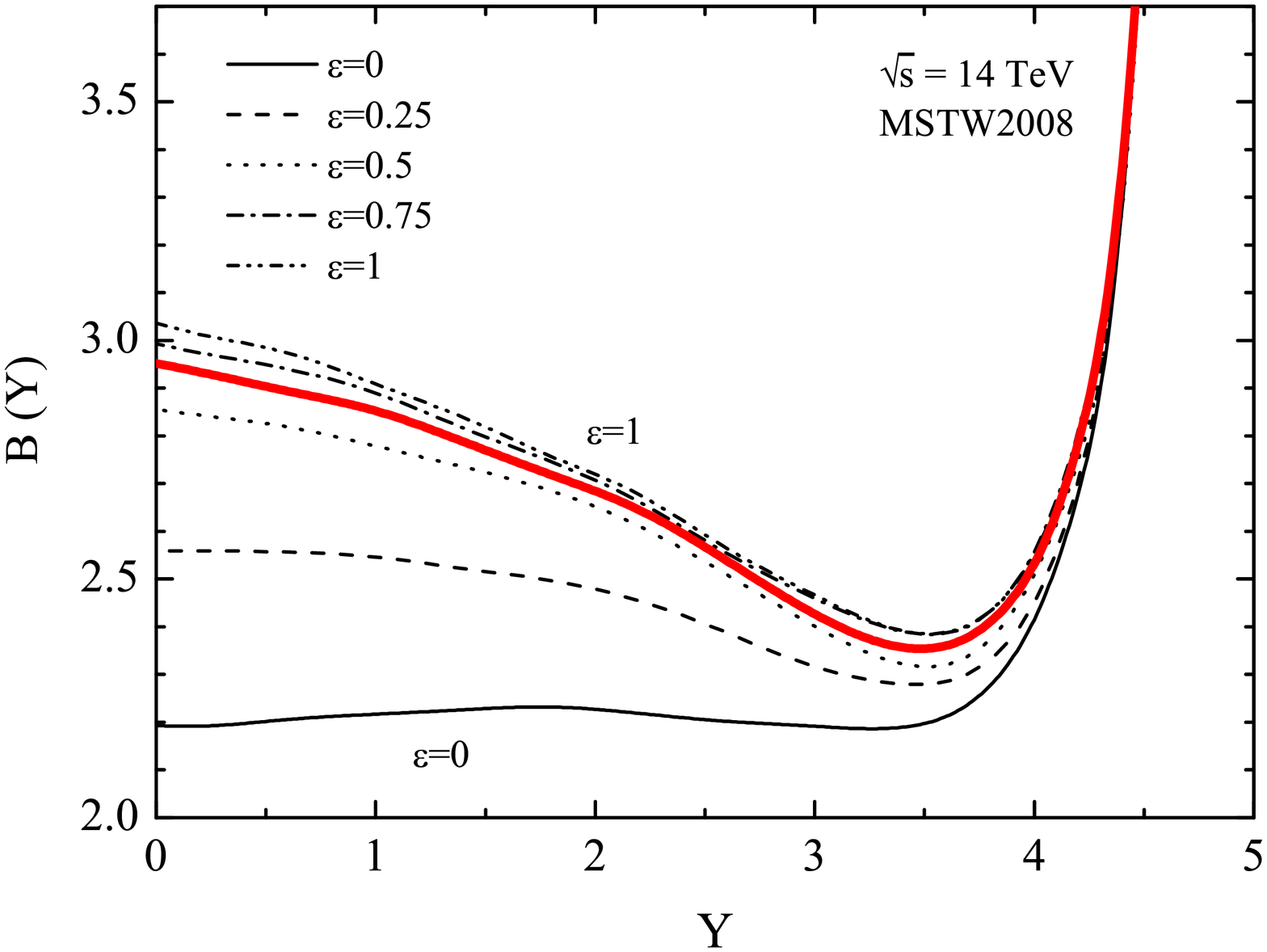}
\caption{
$B(Y)$ as a function of $\epsilon$ ($\epsilon \equiv 2c(x)/(\bar{u}(x)+\bar{d}(x))$)
for $\sqrt{s}$ = 7 and 14 TeV. The CTEQ6.6  and MSTW 2008 PDF sets are used.
The red solid lines show the results including the charm structure function in the sample PDFs.
 }%
\label{fig:by}%
\end{figure}

In Fig. \ref{fig:by}, $B(Y)$, given by Eq. (\ref{eq:By}), is shown
for the several values of the charm contribution parameter $\epsilon$, which is defined as
\be
\epsilon \equiv \frac{c(x)}{((\bar{u}(x)+\bar{d}(x))/2)}
= \frac{2 c(x)}{(\bar{u}(x)+\bar{d}(x))} \ .
\label{eq:eps}
\ee
No cut on the transverse momentum $P_T$ is applied.
The $\epsilon$ = 0 is the case that there is no charm in the proton,
and if the charm contribution is the same as the average of the main sea quarks,
$(\bar{u}(x)+\bar{d}(x))/2$, the parameter $\epsilon$ is 1.
Because the charm is a heavy quark, the charm density would not exceed $\bar{u}(x)$ and $\bar{d}(x)$,
thus, we can probe $\epsilon$ from 0 to 1.
The relative charm contribution $\epsilon$ can be found from the measurement of $B(Y)$.
The result of $B(Y)$ evaluated with the charm structure functions parameterized in the CTEQ6.6
and the MSTW2008 PDFs is presented with the red-solid curve.
As shown in the figure, the value of $\epsilon$ is between 0.5 - 0.75 for both PDF sets.

\begin{figure}[ht]%
\centering
\includegraphics[scale = 0.33]{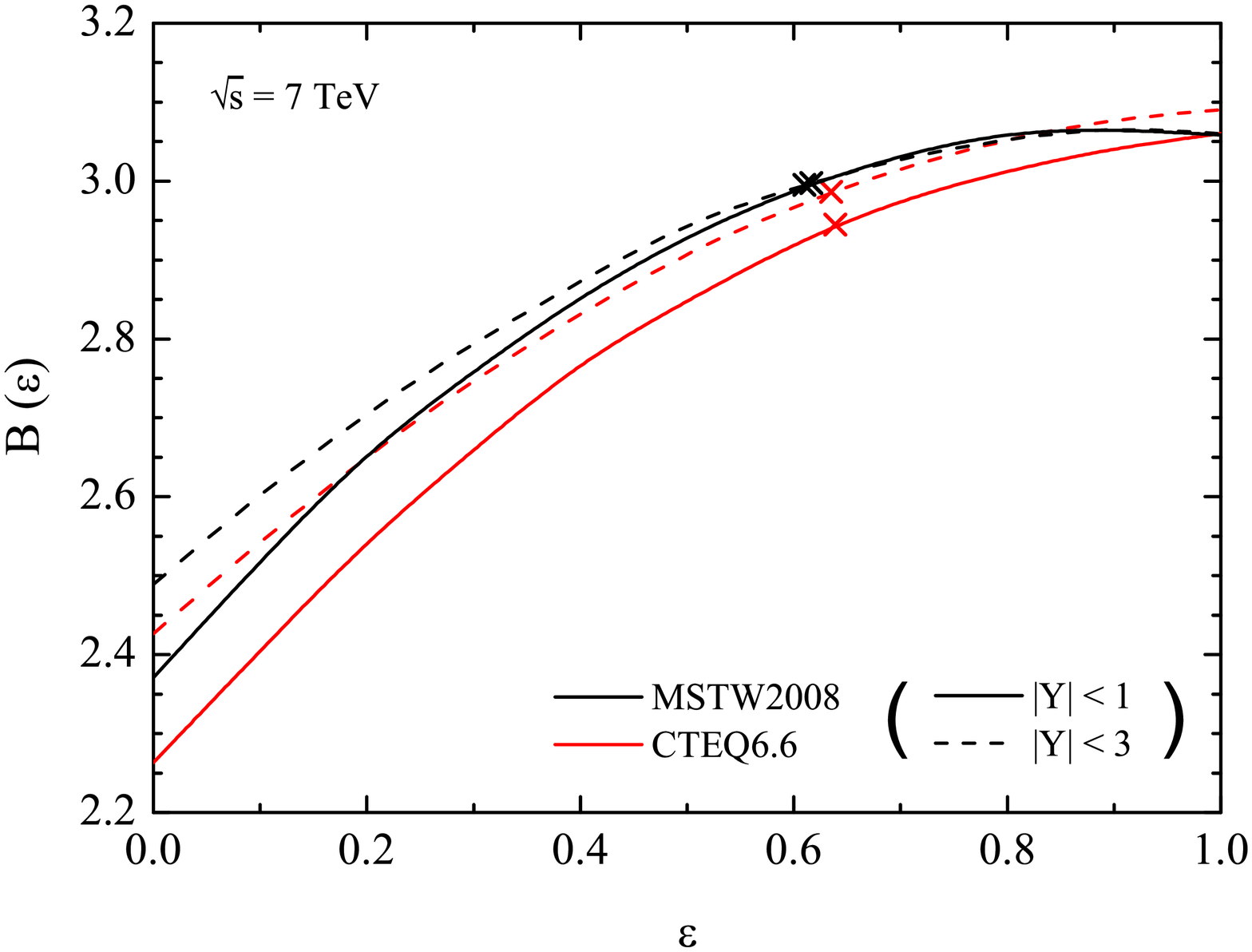}
\includegraphics[scale = 0.33]{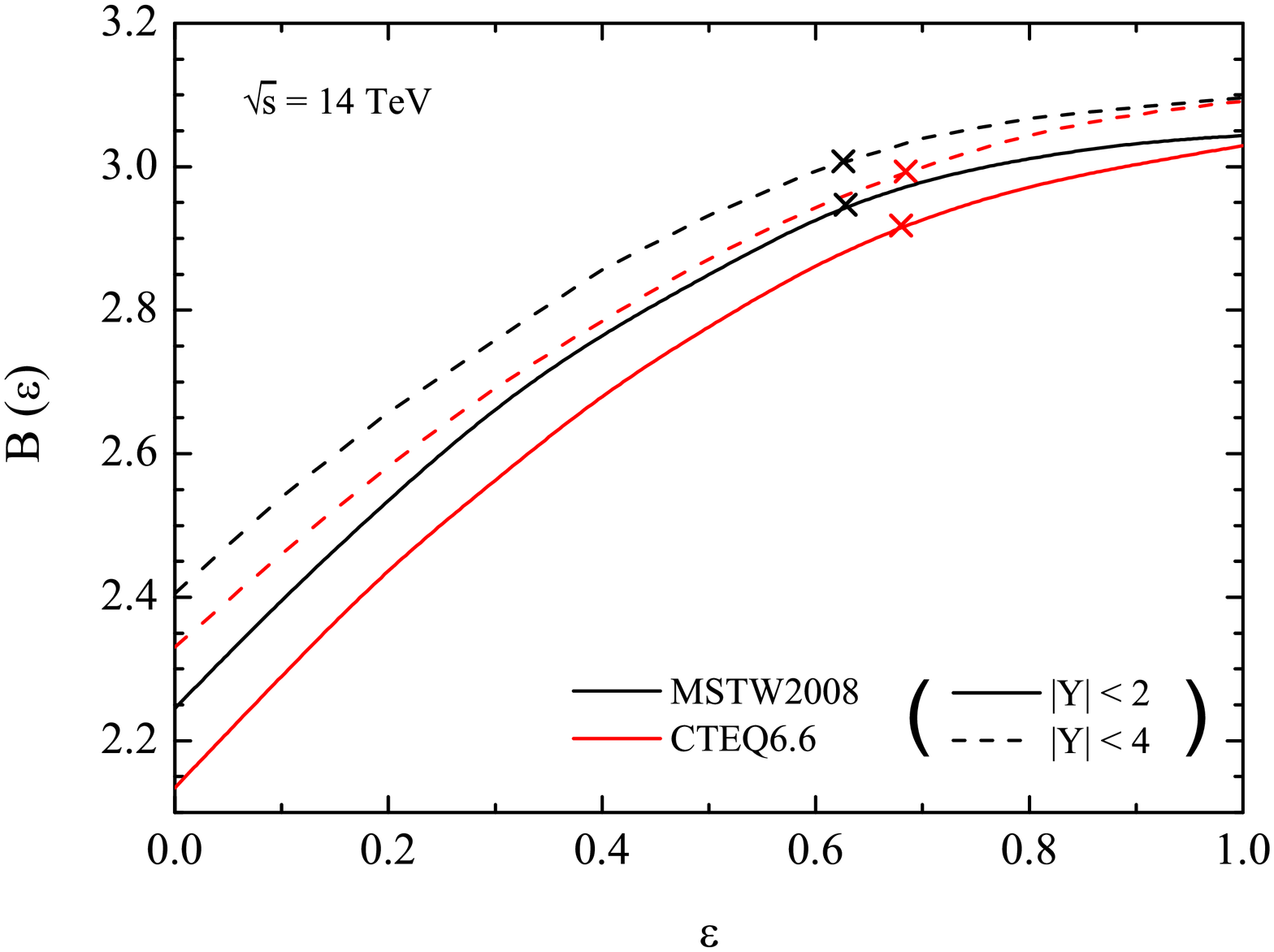}
\caption{The ratio of the integrated cross sections $B(\epsilon)$ as a function of $\epsilon$ for different ranges of Y. The values of $B(\epsilon)$ evaluated with the charm distributions from each sample PDF are represented by the crosses.}%
\label{fig:b-eps}%
\end{figure}

We also calculate the ratio of the integrated cross sections, $B(\epsilon)$,
which is given by
\be
B(\epsilon) = \int_{{\rm with}~Y~{\rm cut}}
\frac{ d\sigma_{W^+}/dY_- + d\sigma_{W^-}/dY_-} {d\sigma_Z/dY} \ .
\label{eq:beps}
\ee
The result is displayed in Fig. \ref{fig:b-eps} as function of $\epsilon$
for several cut-offs on the rapidity, $|Y| \leq$ 1 and 3 for $\sqrt{s}$ = 7 TeV, and 2 and 4 for 14 TeV.
In addition to $B(Y)$, one can determine the $\epsilon$ value from this quantity $B(\epsilon)$.
The crosses in the figure indicate the value of $B(\epsilon)$ evaluated with the charm distributions from the PDF sets.
For example, $B(\epsilon)$ with the cutoff of $|Y|<1$ is 2.95 for CTEQ6.6 \cite{CTEQ66}
and 3.00 for MSTW2008 PDFs \cite{MSTW08} at $\sqrt{s}$ = 7 TeV.
At $\sqrt{s}$ = 14 TeV, it has the value of 2.92 for CTEQ6.6
and 2.95 for MSTW2008 PDFs for $|Y|<2$.
These values yield $\epsilon$ 0.64 to 0.62 at $\sqrt{s}$ = 7 TeV,
and 0.68 to 0.63 at 14 TeV for the CTEQ6.6 and MSTW2008 PDF sets, respectively.


\section{Summary and conclusions}

In this paper, using the sample PDFs, we have examined the charge asymmetry $A(Y)$ for $u$ and $d$ quarks as well as the ratio of cross sections $B(Y)$ and $B(\epsilon)$ in the experimentally accessible weak boson rapidity range at the LHC. We subsequently focused on the role of charm quark contributions.
For the charge asymmetry the effects of the heavy quark PDFs $c(x)$ and $b(x)$ are negligible, even at the highest energy. One can therefore isolate and explore the $u(x)$ and $d(x)$ (or $u_v(x)$ and $d_v(x)$) structure functions.
In this context, we presented a simplified expression of the charge asymmetry in terms of the $u$ and $d$ distribution functions.

We also studied the charm contribution $\epsilon$ = $2c(x)/(\bar{u}(x)+\bar{d}(x))$ to $B(Y)$ and $B(\epsilon)$.
The CTEQ6.6 \cite{CTEQ66} and MSTW2008 PDF \cite{MSTW08} sets indicate a value of $\epsilon$ of about 0.6 for $\sqrt{s}$ = 7 TeV, and 0.6 - 0.7 for 14 TeV.
The measurement of the rapidity distributions of $W/Z$ (or $B(Y)$) interpreted in terms of $B(Y)$ and $B(\epsilon)$ will lead to a straightforward determination of the charm quark component of the sea quarks at the LHC.


\begin{acknowledgments}
The work was supported by the National Research Foundation of Korea (NRF)
grant funded by Korea government of the Ministry of Education, Science and
Technology (MEST) (No. 2011-0017430) and (No. 2011-0020333). The work of F.~H. is supported in part by the National Science Foundation under Grants No. OPP-0236449 and and PHY-0969061, and in part by the University of Wisconsin Alumni Research Foundation.
\end{acknowledgments}

\end{document}